\documentclass[prb,aps,floats,amssymb,showkeys,showpacs,preprint,
superscriptaddress,tightenlines]{revtex4}
\usepackage{graphicx}
\usepackage{amsmath,latexsym,epsf}
\usepackage{epsfig,,amsfonts,bm}
\usepackage{amssymb}
\usepackage{psfrag}
\begin{document}
\preprint{Bicocca-FT-**-**  May 2008}

\title
 {A  study of the $(m,d,N)=(1,3,2)$ Lifshitz point and of the three-\\
dimensional  XY universality class  by high-temperature bivariate\\
 series for the  XY  models  with anisotropic competing interactions\\}
 
\author{P. Butera\cite{pb} }
\affiliation
{Istituto Nazionale di Fisica Nucleare \\
  Sezione di Milano-Bicocca\\
 3 Piazza della Scienza, 20126 Milano, Italy\\}

\author{M. Pernici\cite{mp}}
\affiliation
{Istituto Nazionale di Fisica Nucleare \\
 Sezione di Milano\\
 16 Via Celoria, 20133 Milano, Italy}
\
\date{\today}
\begin{abstract}
High-temperature bivariate expansions have been derived for the
two-spin correlation-function in a variety of classical lattice XY
(planar rotator) models in which spatially isotropic interactions
among first-neighbor spins compete with spatially isotropic or
anisotropic (in particular uniaxial) interactions among
next-to-nearest-neighbor spins.  The expansions, calculated for cubic
lattices of dimension $d=1,2$ and $3$, are expressed in terms of the
two variables $K_1=J_1/kT$ and $K_2=J_2/kT$, where $J_1$ and $J_2$ are
the nearest-neighbor and the next-to-nearest-neighbor exchange
couplings, respectively.  This report deals in particular with the
properties of the $d=3$ uniaxial XY model (ANNNXY model) for which the
bivariate expansions have been computed through the 18-th order, thus
extending by 12 orders the results so far available and making a study
of this model possible over a wide range of values of the competition
parameter $R=J_2/J_1$.  Universality with respect to $R$ on the
critical line separating the para- and the ferro-magnetic phases can
be verified, and at the same time the very accurate determination
$\gamma=1.3177(5)$ and $\nu=0.6726(8)$ of the critical exponents of
the susceptibility and of the correlation-length, in the
three-dimensional XY universality class, can be achieved. 
For the exponents at the multi-critical
$(m,d,N)=(1,3,2)$ Lifshitz point the estimates 
$\gamma_l=1.535(25)$, $\nu_{\bot}=0.805(15)$ and  $\nu_{\|}=0.40(3)$
are obtained.
 Finally, the susceptibility exponent
is estimated along the boundary between the disordered and the
modulated phases.

\end{abstract}

\pacs{ 05.50.+q, 05.70.Jk, 64.60.Kw} 
\keywords{XY model, planar rotator model, ANNNXY model, N-vector model,
 high-temperature expansions, next-nearest-neighbor
interaction, competing spin-spin interactions, spatially anisotropic spin-spin
interactions,  universality, multicritical Lifshitz point, frustrated
spin systems, lattice field theory, Symanzik improvement.}

\maketitle

\section{Introduction} 
 Bivariate high-temperature(HT) and, in some cases,
 low-temperature(LT) series expansions have been derived in the last
 four decades only for very few lattice spin models with interactions
 extending beyond nearest-neighbor($nn$) sites.  In principle,
 undertaking such calculations should be of permanent interest,
 because the analytic approximations based on the series coefficients
 can make a large part of the (at least) bidimensional
 interaction-parameter space easily accessible to analysis above
 (respectively below) the transition temperature.  In practice,
 however, the presence of the next-to-nearest neighbor ($nnn$)
 interactions makes the derivation of adequately long expansions by
 the conventional graph techniques, very laborious. As a consequence,
 only data at relatively low orders and therefore of limited use, have
 been so far available for a handful of non-trivial models and the
 less tedious and more flexible approach to the study of these systems
 by stochastic simulations has largely prevailed in the literature,
 despite its forced limitation to a coarse-grained survey of the
 interaction-parameter space and the convergence problems often met
 with in particular regions of the phase diagrams.

 The earliest studies of short two-variable\cite{Dalton} (or, in some
 cases, even three-variable\cite{Philhours}) series, particularly in
 the case of $N$-vector spin systems, helped to substantiate and
 qualify the critical universality hypothesis in its statement
 concerning the independence of the critical exponents (and of the
 other universal quantities) on the range of the interaction.  
Decisive progress in the study of this property was
 achieved much later with the advent of simulation algorithms
 optimized\cite{Luijte} for long-range interactions. It was however
 already known that, when the $nnn$ interactions compete with the $nn$
 interactions, frustration occurs (in absence of disorder) and, as
 initially shown in the mean-field approximation\cite{Yosh,El},
 produces ``special effects''\cite{SelFi}.  The physically interesting
 new features include :

{\it a)} the formation of spatially modulated phases, i.e.
 spin configurations in which the order parameter varies periodically
 in space with a characteristic modulation wave-vector depending on
 the temperature and the ratio of the competing exchange couplings
 \cite{Yosh,El,SelFi,Step,Hornreich,Hor80,Fith,Selke,Seldl,Yeom,Seu,Lieb,Diehl,Diep,Lois};

{\it b)} the occurrence of a special multi-critical
 point\cite{Hornreich}, called Lifshitz point (LP), at which the HT
 disordered phase meets both the LT spatially-uniform ordered phase
 and the LT spatially-modulated ordered phase(s).

  Competing interactions and LP's are present in a variety of
 magnetic, ferroelectric, polymeric, liquid crystal systems, in
 microemulsion models etc., which sometimes have not yet been studied
 in complete detail
 theoretically\cite{Selke,Seldl,Yeom,Seu,Diehl,Diep,Lois,Malakis,Chit,Wid}
 or experimentally\cite{exp} and continue to be actively explored.
 It is also worth to mention that  within the lattice approach
 to Euclidean quantum field theory there is a continuing interest into
 models of the same or analogous structure\cite{Symanzik,Sheik}, which
 are expected to show a faster approach to the continuum limit.

\section{The spin models}

We begin with a general description of several systems of $N$-vector
spins with $nn$ and $nnn$ interactions on simple-cubic lattices of
spatial dimensions $d=1,2$ and $3$, in zero external field, for which
we have computed from the beginning or extended the bivariate HT
series.  Then we shall discuss a first brief analysis of a small
sample of the large body of data so far accumulated, referring to a
particular three-dimensional system of spins with $N=2$ components,
i.e. to a classical XY (or planar rotator) model.  The case of general
$N$ will be the subject of forthcoming work.

We have derived bivariate HT expansions for:

{\it i)} a class of $d$-dimensional models with isotropic interactions
$J_1$ among $nn$ spins and isotropic (or anisotropic) interactions $J_2$
among $nnn$ spins separated by two lattice spacings along $m \leq d$ lattice
axes. They are described by the Hamiltonian:
\begin{equation}
H_{nnn} \{ v \} = -2{J_1}  \sum_{nn } 
\vec v({\vec r}) \cdot \vec v({\vec r\;'}) - 2{ {J_2} }
\sum_{nnn } 
\vec v({\vec r}) \cdot  \vec v({\vec r\;'} )
\label{tnhamilt}
\end{equation}
 where we have denoted by $\vec v({\vec r})$ a $N$-component classical spin
 vector of unit length situated at the lattice site ${\vec r}$.  In
 eq.(\ref{tnhamilt}) the first sum, extended to $nn$ spins, describes
 the interactions among the spin at $\vec r$ and the spins at the
 sites ${\vec r\;'}={\vec r}+\hat x_i$ with $\hat x_i$ a unit lattice
 vector in the  positive $x_i$ direction.  The second sum describes the
 interactions among the spin at $\vec r$ and the spins at the sites
 ${{\vec r\;'}=\vec r}+2\hat x_i $ with $i=1,..,d$ in the isotropic
 $d-axial$ case (i.e. in which spins separated by two spacings
 interact along all lattice axes), whereas $i$ takes only the values
 $i=d-m+1,..,d$ in the anisotropic $m-axial$ case, with $m<d$.  In
 this report we study in detail only the most interesting case of the
 $m=1$ model in $d=3$, which is sometimes denoted also as the
 three-dimensional ANNNXY model.

{\it ii)} a class of models 
with isotropic or anisotropic interactions $J_1$ among $nn$ spins 
and $J'_2$ among
 (geometric) second-neighbor ($sn$) spins (sometimes also called 
${\it J_1-J_2}$ model or model with crossing bonds) described by the
 following Hamiltonian:
\begin{equation}
H_{sn}\{ v \} = - 2{J_1}  \sum_{nn} 
\vec v({\vec r}) \cdot \vec v(\vec r\;') - 2{ {J'_2} }
\sum_{sn } 
\vec v({\vec r}) \cdot \vec v(\vec r\;')
\label{snhamilt}\end{equation}
The first sum in eq.(\ref{snhamilt}) has the same meaning as in
 eq.(\ref{tnhamilt}), while the second sum extends to $sn$ spins and
 describes the coupling of the spin at $\vec r$ with the spins at the
 sites ${\vec r\;'}={\vec r}+\hat x_i+\hat x_j$ and ${\vec r\;'}={\vec
 r}+\hat x_i-\hat x_j $ with $i<j=1,..,d$ (i.e.  the $sn$ interaction
 acts along the diagonals of the elementary plaquettes).

In eq.(\ref{tnhamilt}) we have denoted by $J_1$ and $J_2$ the $nn$ and
the $nnn$ exchange interaction constants respectively.  We shall
denote by $R=J_2/J_1$ their ratio, which measures the degree of
competition of the couplings and therefore is usually called {\it
competition parameter}.  In eq. (\ref{snhamilt}) $J'_2$ denotes the
$sn$ interaction constant and we set $R'=J'_2/J_1$.  Our HT expansions
are expressed in terms of the two variables $K_1=\beta J_1$ and
$K_2=\beta J_2$ ($K'_2=\beta J'_2$), with $\beta= 1/kT$, $k$ the Boltzmann
 constant and $T$ the temperature.

By the same techniques used in this paper, expansions can be derived
also for a much wider variety of systems, but here we shall not be
concerned with this possibility.

In the case of the $1-axial$ XY model in $d=3$ studied in this report,
our series extend through the 18-th order (altogether 171 nonzero
coefficients in the case of the susceptibility) the 
existing\cite{Redner} sixth-order results (21 nonzero coefficients,
respectively).  The extensions obtained for the other models described
in {\it i)} and {\it ii)} are comparable. 

 The complete set of our series
coefficients is too extensive to print here. We shall however upload in
the hep-lat archive a separate report containing a large sample of
these data, including the correlation functions between $nn$ and $sn$
(or $nnn$) spins, the energy density, the susceptibility and a few
correlation moments for the models we have studied.

It is fair to note that, in spite of their high computational
complexity and cost, our bivariate series still reach a length which
may still be considered  only ``moderate'' with respect to the current
best standards for univariate HT expansions. For example, as far as
the accuracy of the numerical output of their analysis is concerned,
they are not yet comparable with the 25-th order univariate series
computed in the case of three-dimensional scalar spin systems with
$nn$ interactions\cite{bcisiesse} only, (such as the Ising model with
generic spin or the lattice Euclidean scalar field), or with the 26-th
order series derived\cite{bper} in the case of the two-dimensional XY
model with $nn$ interactions on the square lattice. However, if the
comparison is limited to the bivariate series calculated until now for
the systems {\it i)} and {\it ii)} (see below), our expansions seem to
be already non-trivial enough to justify an update of the few existing
HT studies or a first analysis of the data so far unavailable.

 We have studied the expansions of observables defined in terms of
 two-spin correlations, as functions of $K_1$ at fixed values of $R$,
 using, for the moment, only the established {\it single-variable}
 methods\cite{guttmann} of series analysis, namely Pad\`e
 approximants (PA) or inhomogeneous differential approximants (DA), and,
 adopting protocols of analysis well tested in earlier
 papers\cite{Redner,Oitmaa,FerVe,OitVe,Ferer}, we have indeed produced
 reasonably accurate results which can complement or improve those
 from different approaches.  The choice of single variable methods of
 numerical analysis, which might perhaps be considered partly responsible of
 the limited accuracy of our numerical results, was dictated only by
 simplicity, but it is likely that now the HT series are  long enough to
 deserve the additional effort of an analysis by tools more powerful
 and better suited to describe bivariate critical behavior, such as the
 partial-differential\cite{FisherPDA,guttmann} approximants or even
 some simpler two-variable generalization of PAs\cite{Alab}.

For $N>1$, the lower critical dimension of the so called $(m,d,N)$
 Lifshitz point occurring in a $m-axial$ $N$-vector system in $d$
 dimensions, is $d_l(m)=2+m/2$, while the upper critical
 dimension\cite{Hornreich,Seldl} is $d_u(m)=4+m/2$. Therefore one has to
 expect that only in dimension $d>2$, for the $1-axial$ model, a LP
 will show up at $T_{LP} >0$ with a non-classical critical behavior.
 Thus we can use our HT series for the three-dimensional ANNNXY model
 to locate its LP and to obtain estimates of its critical exponents,
 for which some predictions from other approximation methods already
 exist.

 More generally, it is also of interest to analyze the behavior of the
series as functions of $R$ all along the critical line separating the
paramagnetic from the ordered phases. In particular, along  the
critical line between the disordered and the ferromagnetic phases we
can take advantage in our analysis of the old suggestion\cite{ Zinn81,
Fisher} that the accuracy in the determination of universal critical
parameters of a given $N$-vector model, such as critical exponents and
universal amplitude ratios, can be significantly improved by extending
the analysis to a one-parameter family of models belonging to the same
universality class. One should then simply ``tune'' the family parameter 
in order to minimize
or, if possible, to suppress the amplitudes of the (non-universal)
leading non-analytic corrections to scaling in the observables used to
evaluate the critical parameters.  This procedure can be implemented
by a Monte Carlo (MC) method only\cite{Bloete}, by HT-series assisted
MC\cite{xypisa} , or by HT series only\cite{bcisiesse}, and has been
successfully applied in a variety of cases involving $nn$ and local
spin interactions, so far with a single exception\cite{Bloete} for the
spin-$1/2$ Ising model on the simple-cubic lattice.  In this case, a
high-precision MC analysis showed that turning on an isotropic
ferromagnetic coupling beyond $nn$ results into a considerable
decrease of the leading correction-to-scaling amplitudes.
Unfortunately, this simulation study had to be restricted to a single
well-guessed value of the $nnn$ coupling, because ``tuning'' the
additional (irrelevant) interaction to search for its best value, was
too time-consuming.  Now, in the case of the $N-$vector model with
$N=2$, our expansions are sufficiently extended to produce accurate
analytic approximations enabling us to determine in a straightforward 
 way the optimal value of
the irrelevant $nnn$ coupling within a fair approximation.

The LP can be located accurately and the exponents of the
 susceptibility and of the transverse correlation-length can be
 determined with fair precision, while the exponent of the parallel
 correlation-length can  be somewhat less accurately estimated. We have
 also examined the critical behavior of the susceptibility along the
 branch of the critical line between the disordered and the modulated
 phases.  In this case the results are significantly 
 less accurate and somewhat
 puzzling: in particular it is not clear whether the transition is 
(weakly) first-order or second-order and, if this is the case,
 to which universality class it belongs. 

Finally, we can point out another valuable use of our expansions as a
 guide for possibly more detailed MC investigations of models in the
 classes {\it i)} and {\it ii)}, which might focus on specific points
 of the parameter space. Even more simply, our expansions can also serve as
 a realistic test-ground for techniques of analysis and re-summation
 of multivariate series.  If nothing else, our calculations provide,
 for a variety of models, an initial set of HT reference data large
 enough to be a stringent constraint in the validation of future
 series extensions.

For completeness and in order to put our work into perspective, it is
convenient to list the main existing HT results for the systems {\it
i)} and {\it ii)} and  to mention a few studies related to the
subject of this paper, but using different techniques.

The earliest bivariate HT series investigation\cite{Redner} of the LP in
 the $N$-vector model with $nn$ and uniaxial $nnn$ interactions, was
 mainly devoted to the $N=1$ case (namely the spin-1/2 Ising model).
 In three dimensions, series of order $8,6$ and $5$ were derived for
 $N=1,2$ and $3$, respectively. However, only the Ising HT series were
 considered\cite{Redner} by the authors long enough to yield
 sufficiently reliable numerical estimates of the location and the
 exponents of the LP.  Both in two and three dimensions, the $N=1$
 series were subsequently extended\cite{Oitmaa,FerVe,OitVe} through
 order 11 for the ${\it J_1-J_2}$ interaction, for the {\it $1$-axial}
 and for the {\it $d$-axial} interactions.  Soon later, series for the
 susceptibility through the twelfth order were computed and
 analyzed\cite{Ferer} for uniaxial models on the simple-cubic and the
 face-centered-cubic lattices.  More recently, in two dimensions for
 the square-lattice Ising model, the susceptibility expansion was
 pushed\cite{Soe} through the thirteenth order for the ${\it J_1-J_2}$ and
 the {\it 2-axial} interactions.  These have been until now the
 longest bivariate series derived for systems with $nn$ and $nnn$
 interactions.

 For the $N \geq 2$ models, on the other hand, there has been no
 further progress in the HT calculations for the last three decades
 since the early sixth-order results\cite{Redner} in three dimensions.
 The only exception to this lack of activity was a study of the
  {\it 2-axial} $N$-vector model (with isotropic $nn$ and $nnn$ 
interactions on the square lattice), in which the bivariate HT series 
expansion coefficients were computed and tabulated through
 the fifth-order\cite{Clarizia}, for {\it general} values of $N$.
 Interest into this model, which constitutes the ``Symanzik improved''
 \cite{Symanzik} square-lattice formulation of the $O(N)$-symmetric
 non-linear $\sigma$-model in two-dimensional Euclidean field theory,
 came however from quantum field theorists. In particular, these HT
 expansions were derived as a means to infer the weak-coupling
 (i.e. LT) properties of the ``improved'' non-linear $\sigma$-model 
for $N \geq 3$.

  A variety of MC investigations of models with $nnn$ interactions can also
 be found in the literature: most of them are concerned with the $N=1$
 case\cite{Binder,SelFi,Sel78,Sel79,Kaski,Ferer,FerVe,PHENK}, and 
 some, more
 directly related to our work, with the $N=2$ case.  In
 particular, the uniaxial XY model in three dimensions was simulated in
 Refs.[\onlinecite{Selke78,Selke80}], its phase diagram was 
 sketched out and the location of the LP was estimated along with the
 exponents of the susceptibility and the magnetization. Later 
 investigations\cite{Ferna,LoisonSimon} were devoted also to the
 $J_1-J_2$ XY model in two dimensions.  
These studies were generally limited to only
 a few points in the interaction-parameter space.

Before a systematic renormalization group approach could be extended
to cover also the class of systems studied here, non-trivial technical
difficulties, due to the anisotropy of the scale invariance in the
critical behavior at LP's, had to be solved.  In particular, while the
lowest order computation of all critical exponents by the
$\epsilon$-expansion around the LP upper critical dimension $d_u(m)$
(with $\epsilon=d_u(m)-d$), as well as some $O(\epsilon^2)$ results, go
back to three decades ago\cite{Hornreich}, the extension through
second order has been completed\cite{DiSh,lemer} only recently.  It is
encouraging that the $O(\epsilon^2)$ corrections are small and
decrease with increasing $N$, so that a simple truncation to order
$\epsilon^2$ of the exponent expansions might already lead to
reasonable approximations.  Of course, further work is still needed
before an accuracy comparable to that established for the usual
critical points can be attained.

  Results consistent with the
$O(\epsilon^2)$ calculations for the exponents of the Ising $1-axial$
model in three dimensions have also been obtained\cite{Berv} from a
truncation of the exact renormalization group equation.
 Renormalization group discussions of the ANNNXY model have been given 
in Refs.[\onlinecite{Garel,ahamuk}].

  There has been recent progress\cite{Shpot} also in the calculation
of the leading non-trivial corrections to the large $N$
limit\cite{Hor} for some exponents.

The layout of the paper is the following: In Sect. II the quantities
for which we have computed HT expansions are defined in detail, and
related to the main critical parameters in particular to the exponents
of the uniaxial LP.  The Sect. III is devoted to a discussion of the
numerical analysis of the series. In the Appendix, we briefly describe
the non-graphical algorithm used to compute the HT expansions and the
checks passed by our series data.

\section{The high-temperature expansions}

In zero field and for any dimension $d$, for all models of the classes
 {\it i)} and {\it ii)} defined on bipartite lattices, the free energy
 is an even function of $J_1 $ and therefore we can restrict our
 analyses to the case $J_1 > 0 $ (ferromagnetic $nn$ interaction).

In general, for $d>1$ these models exhibit three main phases: a HT
paramagnetic ({\it P}) phase, a LT uniformly-ordered ferromagnetic
({\it F}) phase and a family of LT ordered  modulated ({\it M}) phases.
For $d>1 $, a non-trivial transition line $K_{1c}(R)$ ($K_{1c}(R')$)
separates the HT {\it P}-phase from the LT phases, whereas for $d=1$,
one has simply $K_{1c}(R)=\infty$.  For $d>d_l(m)$, the critical line
$K_{1c}(R)$ is divided into two branches by a LP, a triple point
located at a non-zero value $K_{1c}(R_{LP})$ and separating the {\it
P-F}-transition line from the {\it P-M} line.

For $T=0$, the spin ordering can be determined simply by minimizing the
energy. The ferromagnetic ground state is energetically favored over
the modulated phase(s) only for $R$ greater than some critical value.
In the three-dimensional uniaxial case under study, the ground state
is made of ferromagnetic layers orthogonal to the $z$-axis, with the
relative orientation of the successive layers determined by the value
of $R$.

It is clear that, except in the $d=1$ case, in which the LT region is
 simply shrunk to the border of the paramagnetic region, the zero-field HT
 expansions are unsuited to yield much more than hints on the LT
 structure of the phase diagram and therefore, for the purpose of
 investigating the LT phases, they have to be replaced by other methods.

For all models of the classes {\it i)} and {\it ii)}, we have derived
the bivariate expansion of the spin-spin correlation-function,
\begin{equation}
C(\vec 0, \vec x;K_1,R)= <{\vec v}(\vec 0) \cdot {\vec v} (\vec x)>,
\label{corfun}
\end{equation}  
 for all values of $\vec x$ for which non-vanishing coefficients exist
within the maximum order of expansion. 

The appropriate quantities to be studied in order to locate
 the {\it P-F}  
branch of the critical line are the ordinary susceptibility
\begin{equation}
\chi(K_1,R )= 1+ \sum_{\vec x \neq 0 } 
<{\vec v}(\vec 0) \cdot {\vec v}(\vec x)>   
\label{ordsus}
\end{equation}
and  the  $l$-th order 
spherical moments   of the correlation function
\begin{equation}
{\rm m}^{(l)}(K_1,R) = 
\sum_{\vec x  }|\vec x|^l <{\vec v}(\vec 0) \cdot {\vec v}(\vec x)>.
\label{sfermom}
\end{equation}
In terms of ${\rm m}^{(2)}(K_1,R)$ and $\chi(K_1,R )$,  
we can construct the  correlation length 
\begin{equation}
\xi^2(K_1,R) ={\rm  m}^{(2)}(K_1,R)/2d\chi(K_1,R)
\label{corleng}
\end{equation}
When studying a $m-axial$ model with $m<d$ and therefore with
anisotropic interactions, it is convenient to break the
$d-$dimensional lattice vectors ${\vec x}$ as $\vec x = ({\vec
x}_{\|},{\vec x}_{\bot})$, where ${\vec x}_{\|}$ and ${\vec x}_{\bot}$
denote the $m$-dimensional and the $(d-m)$-dimensional components of
${\vec x}$, respectively parallel and perpendicular to the directions
of the $nnn$ interaction.

 In order to study the properties of the LP and to locate the {\it P-M} 
 transition from the paramagnetic to the LT modulated phase it is
 necessary to study also the structure function with respect to $\vec q_{\|}$

\begin{equation}
\chi(\vec q_{\|};K_1,R )=
 1+ \sum_{\vec x \neq 0}  e ^{ i \vec q_{\|} \cdot \vec x} 
<{\vec v}(\vec 0) \cdot {\vec v}(\vec x)>   
\label{structfunpar}
\end{equation}

and to $\vec q_{\bot}$
\begin{equation}
\chi(\vec q_{\bot};K_1,R )=
 1+ \sum_{\vec x \neq 0 }e ^{ i \vec q_{\bot} \cdot \vec x} 
<{\vec v}(\vec 0) \cdot {\vec v}(\vec x)>.   
\label{structfuntrasv}
\end{equation}
  In a vicinity of the LP, the scaling behavior of the two-spin
correlation function becomes strongly anisotropic. In particular, the
asymptotic behavior for large separation of spins joined by a vector
whose components lie entirely in the $m-$dimensional ${\vec x}_{\|}$
subspace, differs from the behavior in the case in which the spins are
joined by a vector in the $(d-m)$-dimensional ${\vec x}_{\bot}$
subspace. It is then necessary to replace\cite{Hornreich} each one of
the usual correlation exponents $\eta$ and $\nu$ by a pair of
exponents associated to the two subspaces.  In the first case, the
{\it parallel} correlation-exponent is usually denoted by $\eta_{l4}$
(or less frequently, but more suggestively by $\eta_{\|}$), in the
second case the {\it transverse} correlation-exponent is denoted by
$\eta_{l2}$ (or by $\eta_{\bot}$). Correspondingly, in different
directions two distinct correlation lengths $\xi_{\|}$ and
$\xi_{\bot}$, are observed which diverge with different exponents
$\nu_{\|}$ and $\nu_{\bot}$, respectively. These exponents are related
to the susceptibility exponent $\gamma_{l}$ at the LP, by the
``anisotropic scaling laws''
\begin{equation}
\gamma_{l}= (2-\eta_{\bot})\nu_{\bot} = (4-\eta_{\|})\nu_{\|}.
\label{aniscal}
\end{equation}
Moreover a generalized hyperscaling law is expected to hold: 
\begin{equation}
2-\alpha_{l} =
m\nu_{\|}+(d-m)\nu_{\bot}.  
\label{iperscal}
\end{equation}
The other scaling relations:
$\alpha_{l}+2\beta_{l}+\gamma_{l}=2$ and
$\gamma_{l}=\beta_{l}(\delta_{l}-1) $ remain
unchanged\cite{Hornreich}.  We can thus conclude that three 
independent exponents are requested to characterize the uniaxial LP.

 When $d>d_u(m)=4+m/2$, and $m=1$, the
critical exponents assume the $N-$independent mean-field values
reported in Table \ref{tab1}.
  
  In order to estimate the additional critical exponents
characterizing a LP, 
we have also computed the $l$-th order ``parallel moments'' of the correlation
function
\begin{equation}
{\rm m}_{\|}^{(l)}(K_1,R)= 
\sum_{\vec x  }|{\vec x}_{\|}|^l <{\vec v}(\vec 0) \cdot {\vec v}(\vec x)>
\label{xparmom}
\end{equation}
and the $l$-th order `` perpendicular moments''
\begin{equation}
{\rm m}_{\bot}^{(l)}(K_1,R)= 
\sum_{\vec x  }| {\vec x}_{\bot}|^l <{\vec v}(\vec 0) \cdot {\vec v}(\vec x)>.
\label{xperpmom}
\end{equation}

Near the LP, the correlation length $\xi^2_{\|}(K_1,R)$ in a direction
 within the $m$-dimensional subspace of the $nnn$ interaction is then
 expressed in terms of these quantities by
\begin{equation}
\xi^2_{\|}(K_1,R_{LP})= 
{\rm m}_{\|}^{(2)}(K_1,R_{LP})/2m\chi(0;K_1,R_{LP})
\sim \tau(R_{LP})^{-2\nu_{\|}}
\label{xipar}
\end{equation}
 and analogously, the correlation length $\xi^2_{\bot}(K_1,R)$ 
 in a direction  orthogonal to the $nnn$ interaction  by 
\begin{equation}
\xi^2_{\bot}(K_1,R_{LP})= 
{\rm m}_{\bot}^{(2)}(K_1,R_{LP})/2(d-m)\chi(0;K_1,R_{LP})
\sim \tau(R_{LP})^{-2\nu_{\bot}}
\label{xicrit}
\end{equation}
 with $\tau(R)=T/T_c(R)-1$.  Defining, in analogy with $\tau(R_{LP})$,
 the reduced competition ratio $ \rho_{LP}=R/R_{LP}-1$, a crossover
 exponent $\phi$ can be introduced to characterize the behavior of the
 reduced critical temperature $\tau_{LP}=T_c(R)/T_c(R_{LP})-1$ as the
 LP is approached along the critical line: $\tau_{LP} \sim
 |\rho_{LP}|^{1/\phi}$ Beside the above exponents, a related\cite{HB} one:
 $\beta_q=\nu_{\|}/\phi$ is associated to the LP.  It describes the
 behavior of the magnitude of the modulation vector $\vec q$ as the LP
 is approached along the branch of the critical line separating the
 disordered and the modulated ordered phases: $|\vec q\;^2| \sim
 \rho_{LP}^{2\beta_q}$.

\section{ The three-dimensional  uniaxial (ANNNXY) model}

\subsection{  Universality along the {\it P-F} branch of the critical line}

For $R$ in the range $-1/4 <R < \infty $, the ground state of the
system is ferromagnetic and the uniform ferromagnetic ordering
persists at $T>0$ up to some inverse temperature $K_{1c}(R)$, at which
a second-order phase transition, expected to belong to the
universality class of the three-dimensional XY model, occurs between
the LT phase and the HT paramagnetic phase.

In this subsection we shall mainly discuss the numerical evidence
 obtained from the analysis of our HT series that, in the
 ferromagnetic range of $R$, this transition actually belongs to the
 XY universality class.  We shall then argue that, if this is the case, the
 parameter $R$ can be exploited to determine very accurate values of
 the critical exponents for the XY universality class.

For each value of $R$, we can locate the transition by analyzing the HT
expansion of the ordinary susceptibility $\chi(K_1,R)$, whose
coefficients show generally a smooth dependence on the order of
expansion and a fast approach to their asymptotic forms, and thus are
well suited to numerical study.

 The critical behavior of the susceptibility as $\tau(R) \rightarrow 0^+$ 
is expected to be
 \begin{equation}
 \chi(K_1,R) =A^{+}_{\chi}(R) \tau(R)^{-\gamma(R)}\Big(1+
 a^+_{\chi}(R)\tau(R)^{\omega(R)}+O(\tau(R))\Big)
\label{chias}
\end{equation}
where $A^{+}_{\chi}(R)$ is the critical amplitude of the
 susceptibility, and $a^+_{\chi}(R)$ is the leading
 correction-to-scaling amplitude. We have denoted
  $\gamma(R)$ and 
 $\omega(R)$ as {\it a priori} $R$-dependent, although we shall
 finally argue that, as expected, they are universal with respect to $R$,
  namely independent of $R$.

The critical behavior of the second-moment 
correlation length  may analogously be characterized as
 \begin{equation}
 \xi^2(K_1,R) =A^{+}_{\xi^2}(R) \tau(R)^{-2\nu(R)}\Big(1+
 a^+_{\xi^2}(R)\tau(R)^{\omega(R)}+O(\tau(R))\Big).
\label{xi2as}
\end{equation}

\begin{figure}[tbp]
\begin{center}
\leavevmode
\includegraphics[width=3.37 in]{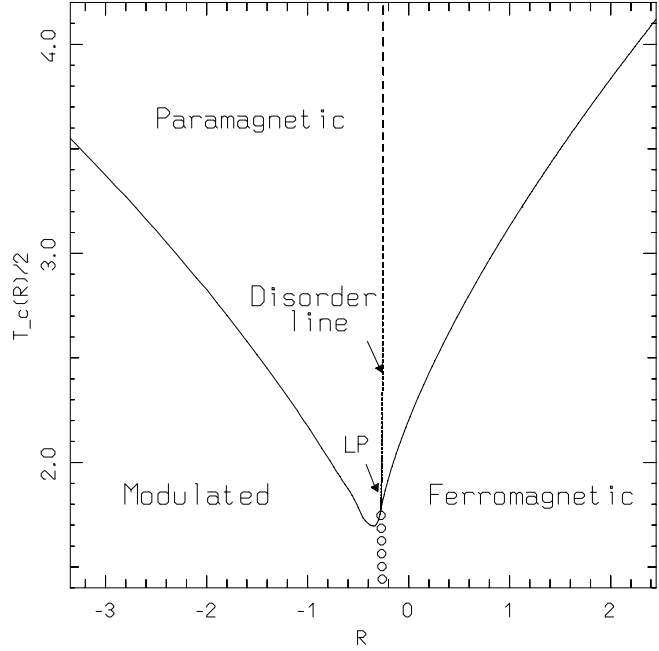}
\caption{ \label{bcrfm} The phase diagram of the three-dimensional
 {\it uniaxial} XY model. In the $(R, T_c(R))$ plane, we have
 represented by a continuous line the locus of critical points
 separating the disordered (paramagnetic) phase from the ordered
 (ferromagnetic and modulated) phases.  In the scale of the figure,
 the uncertainties of the points are smaller than the width of the
 line.  The Lifshitz point (LP) is located at the intersection of the
 critical and the {\it disorder} line (represented by a dashed curve).
 A transition line, also expected to be of second order, separates the
 modulated from the ferromagnetic phase, and joins the Lifshitz point
 to the point $(R=-1/4,T=0)$, where the ordering of the ground state
 changes. This line cannot be mapped out by HT methods and therefore
 is only schematically indicated by a sequence of open circles.}
\end{center}
\end{figure}

We shall re-sum the susceptibility  
 series  by inhomogeneous
 second-order DAs in the variable $K_1$ at fixed $R$.  In this and in
 the analyses that follow, we have used a set of quasi-diagonal DAs
 chosen as the approximants $[k,l,m;n]$ with $14 \leq k+l+m+n \leq
 16$, namely those using not less than 17 series coefficients. We have
 taken $|k-l|$,$|l-m|$, $|k-m|$ $< 3$ with $k,l,m >3$, while $1 \leq n
 \leq 4$. We have however always made sure that our numerical estimates
 depend only  weakly on this choice.  The results for the critical
 line obtained in this way (for $R_{LP} \lesssim R \lesssim 2.$) are
 reported in Fig. \ref{bcrfm}.  Notice that we have preferred to plot
 vs $R$ the quantity $T_c(R)/2=1/2K_{1c}(R)$ rather than $K_{1c}(R)$
 itself, in order to make our figure immediately comparable with the
 figure, covering a smaller range of $R$, which appears in the MC
 study of Ref.[\onlinecite{Selke78}]. On the scale of
 Fig. \ref{bcrfm}, the three data points $(R=0.,T_c(R)/2=2.17(2))$,
 $(R=-0.25,T_c(R)/2=1.83(2))$ and $(R=-0.26,T_c(R)/2=1.82(2))$
 obtained\cite{Selke78} in the MC study, are hardly distinguishable
 from our curve.  The corresponding values determined by our series
 are: $(R=0.,T_c(R)/2=2.2017(2))$, $(R=-0.25,T_c(R)/2=1.830(1))$ and
 $(R=-0.26,T_c(R)/2=1.809(1))$.  Our result at $R=0.$ compares well
 with the estimate $(R=0.,T_c(R)/2=2.20172(15))$, obtained\cite{bcoen}
 by 21th-order HT expansions.  Notice also that the spreads $\delta
 K_{1c}(R)$ of our DA estimates of the critical temperatures, which
 are smaller than those of the MC study by one order of magnitude, are
 invisible on the scale of the figure. For convenience, we have listed
 in Table \ref{tabk} a few numerical values of $K_{1c}(R)$.

It should be stressed that it is generally difficult to assess very
 accurately the real uncertainties of the results in this kind of
 analysis mainly because, due to the finite (and in our case still
 moderate) length of the series, the sequences of DA estimates
 may retain residual trends which call for further
 extrapolation, particularly so for $R \gtrsim 1$.  Whenever possible,
 one should then try to infer the size of the uncertainties also from
 a comparison with the results of approximation procedures alternative
 to the direct DA calculation, and thus presumably having different
 convergence rates and different mechanisms of error build-up.  In our
 case, we shall eventually argue that the spread of the DA estimates
 of the critical temperature, already at these orders of HT expansion,
 are reasonable approximations of the uncertainties, at least for
 positive and not too large $R$.
 
Let us then consider an example of an alternative approach to the
determination of $ K_{1c}(R)$.  If the critical singularity is the
nearest singularity, we can determine it also by evaluating the limit
of the sequence of estimators $\big(K_{1c}(R)\big)_n$ of $K_{1c}(R)$
defined by the modified-ratio prescription\cite{Zinn81,guttmann}:
\begin{equation}
\big(K_{1c}(R)\big)_n=(\frac{ c_{n-2}c_{n-3}} {c_{n}c_{n-1}})^{1/4} 
\exp[\frac{ s_{n}+s_{n-2}} {2s_{n}(s_{n}-s_{n-2})}]
\label{zinnbc}
\end{equation}
where
\begin{equation}
s_{n}=\Big( {\rm ln}( \frac {c_{n-2}^2} {c_{n}c_{n-4}})^{-1}+
 {\rm ln}( \frac {c_{n-3}^2} {c_{n-1}c_{n-5}})^{-1}\Big)/2 
\end{equation}
 and $c_n(R)$ are the HT expansion coefficients of the susceptibility.

 This prescription has the important advantage of bringing
information not only about $K_{1c}(R)$, but also on the leading
correction-to-scaling amplitude $ a^+_{\chi}(R)$, defined by
eq.(\ref{chias}), a quantity which in general rules the convergence
properties of any approximation method in the critical region. Indeed,
we have observed\cite{bcisiesse} that
  the modified-ratio sequence has the simple
asymptotic behavior for large order $n$
 \begin{equation}
(K_{1c}(R))_n=K_{1c}(R)\Big(1-\frac{ C(\gamma,\omega) a^+_{\chi}(R)}
 {n^{1+\omega}}+O(1/n^2)\Big)
\label{zinnkc}
\end{equation}
 where $C(\gamma,\omega)$ is some known positive function of $\gamma$
 and $\omega$.

\begin{figure}[tbp]
\begin{center}
\leavevmode
\includegraphics[width=3.37 in]{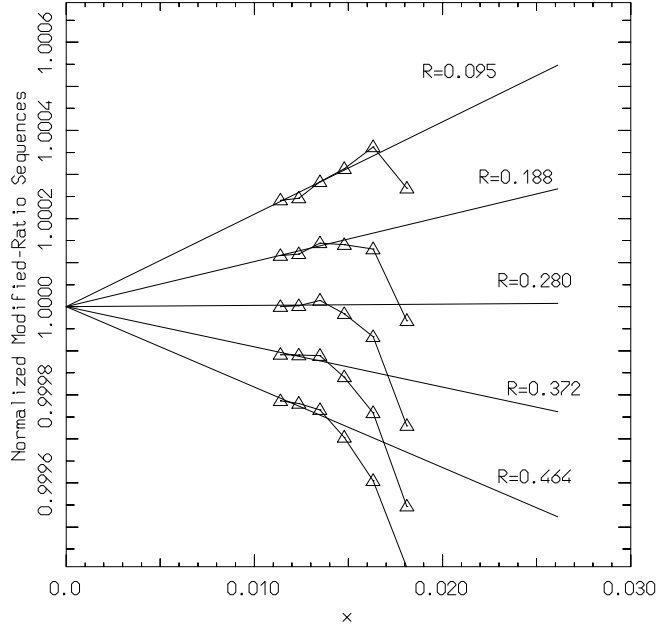}
\caption{ \label{bcmodrat}  
A representation of the modified-ratio sequences $(K_{1c}(R))_n$
obtained from eq.(\ref{zinnbc}).  For a few values of $R$ chosen in a
vicinity of $R_M$, the sequences have been extrapolated to large order
by fitting them to the asymptotic form $ b_1(R)- b_2(R)/n^{1+\omega}$
with $\omega=0.52$.  We have then plotted the modified-ratio sequences
vs $x=1/n^{1+\omega}$, after normalizing them to their limiting values
$b_1(R)$.  The elements of the normalized sequences are represented by
open triangles and are connected by continuous lines to guide the eye.
The corresponding best fits to the asymptotic form: $1.-
a(R)/n^{1+\omega}$ with $a(R)=b_2(R)/b_1(R)$ are represented by
straight continuous lines.}
\end{center}
\end{figure}

  In order to use effectively the modified-ratio method, we shall now
 {\it assume} that the exponent $\omega=\omega(R)$ of the leading
 correction to scaling is independent of $R$, and takes the
 value\cite{bcoen} $\omega \simeq 0.52$ (an assumption which was not
 necessary to make in the DA method discussion).  Then by fitting the sequence
 $\big(K_{1c}(R)\big)_n$ to the simple form $
 b_1(R)-b_2(R)/n^{1+\omega}$, we can estimate the amplitude
 $a^+_{\chi}(R)$ from the value of the parameter $b_2(R)$. Although
 rather long and smooth series are  usually  necessary\cite{bcisiesse}
 to obtain accurate estimates by this method, we can observe a
 complete consistency between the estimates of $K_{1c}(R)$ from the
 DAs and the values of $b_1(R)$ obtained by the fit  (within a small multiple
 of their spreads, for positive and not too large $R$), so that the
 results cannot be distinguished from those reported in
 Fig.\ref{bcrfm}.  This fact also gives support to our simple fit
 procedure for determining $b_2(R)$ and, at the same time, it suggests
 that the spread of the DA estimates of the critical temperatures
 might be taken as a sound measure of their real uncertainties.  In
 order to illustrate our fit procedure for determining $b_2(R)$, in
 Fig.\ref{bcmodrat} we have plotted vs $x=1/n^{1+\omega}$ the sequence
 of modified ratios $(K_{1c}(R))_n$ normalized to their extrapolated
 values $b_1(R)$, for a few fixed values of $R$, chosen in a vicinity
 of $R_M$.  Let us add that only for $R \gtrsim 0$ the sequences
 $(K_{1c}(R))_n$ are sufficiently smooth that these  rough, but
 sufficiently reliable, estimates of $b_2(R)$ are feasible, while
 unfortunately for $R<0$ the modified-ratio sequences develop strong
 oscillations and a straightforward two-parameter fit cannot work.
 The important observation is now that the function $b_2(R)$ vanishes
 at $R=R_M \simeq 0.28(3)$ and therefore its absolute value is minimum
 at this point.  From our estimates of $b_2(R)$, we can infer the sign
 and size of the deviations from the exact values which should be
 expected for the central DA estimates of $K_{1c}(R)$ and of the
 exponents, as $R$ varies.
  More precisely, we have to expect\cite{LiFi} that, in the
 range $R>R_M$, where $b_2(R)$ and therefore $a^+_{\chi}(R)$ are found
 to be positive, the critical inverse temperatures and the critical
 exponents will be underestimated by our analyses, while the opposite
 will be observed for $R<R_M$. This is clear from eq.\ref{zinnkc}, as far as
the critical inverse temperatures are concerned. 
 In order to  reach the same conclusion for the exponents,
 one may either use an asymptotic formula\cite{bcisiesse}  analogous 
to eq.\ref{zinnkc}, or consider that in approximate calculations
  some effective exponent\cite{LiFi} is generally evaluated, 
for example: 
$\gamma_{eff}(\bar \tau)=-\frac {d\rm log(\chi)} {d \rm log(\tau)} 
\simeq \gamma- \omega a^+_{\chi}\bar \tau^{\omega}$ with $\bar \tau$ small,
 but nonzero.
 In order to relate these remarks to the behavior of $b_2(R)$, the
 absolute value of $b_2(R)$, obtained from a fit of the four
 highest-order estimators $(K_{1c}(R))_n$ in the modified-ratio
 sequence, is very schematically plotted vs $R$, together with the
 exponent estimates, in Figure \ref{esprfm} and in some of the
 following figures. All previous considerations also apply to the 
study of the HT expansion of $\xi^2(K_1,R)$.

In Figure \ref{esprfm}, the critical exponents $\gamma(R)$ of the
susceptibility and $\nu_{\bot}(R)$ of the transverse
correlation-length (notice that $\nu_{\bot}(R)$ and $ \nu_{\|}(R)$
coincide with $\nu(R)$ for $R \gtrsim 0$) are plotted as functions of
$R$ along the {\it P-F} branch of the critical line.  They have been
computed both using second-order DAs biased with the critical
singularity $K_{1c}(R)$ and, alternatively, also by the method of
``critical-point renormalization'' \cite{cpr}. The latter method
consists in analyzing the term-by-term divided series $ W(x,R)=
\sum_{s} d_s(R)/c_s(R)x^s$, where $d_s(R)$ are the expansion
coefficients of $\chi^2(K_{1},R)$ (or of $\xi^4(K_{1},R)/K_{1}^2$) and
$c_s(R)$ are the coefficients of $\chi(K_{1},R)$ (or of
$\xi^2(K_{1},R)/K_{1}$).  It can be shown that, if the nearest
singularity is the critical point, then $ W(x,R)$ is singular at $x=1$
with an exponent $-(1+\gamma(R))$ (resp.$-(1+2\nu(R))$), which can be
estimated by forming DAs of $ W(x,R)$ biased to be singular at
$x=1$. This alternative determination of the exponents is of
particular interest in the ranges of values of $R$ in which the
accuracy of the available estimates of $K_{1c}(R)$ might be
insufficient to obtain good temperature-biased estimates or where the
convergence of the DAs is slow.  It is not surprising that in a
neighborhood of $R_M$ (where the amplitude of the leading correction
to scaling is vanishing), this method and the direct analysis of the
susceptibility (or of the correlation length) by temperature-biased
DAs yield essentially the same exponent estimates, while elsewhere
they show some small difference. In Fig. \ref{esprfm}, the results of
both approximations are plotted vs $R$ and compared with a recent high
accuracy determination\cite{xypisa} $\gamma=1.3178(2)$ and
$\nu=0.67155(27)$ of the exponents $\gamma$ and $\nu$ for the XY
universality class.  Our two approximations for $\gamma(R)$ and
$\nu(R)$ show a similar behavior: for $R>R_M$ both lead to estimates
slightly smaller than the data chosen for comparison, while the
opposite happens for $R<R_M$. This is precisely what can be
anticipated from our determination of the leading
correction-to-scaling amplitude.  Overall, as shown in Figure
\ref{esprfm}, for $0.1 \lesssim R \lesssim 1.$, the central values of
our estimates of the critical exponents deviate from the estimates at
$R=R_M$ by less than $0.5\%$, while in the wider range $0. \lesssim R
\lesssim 1.5$ the deviations do not exceed $1\%$, thus indicating that
our approximate results have a very weak dependence on $R$, to within
a fair accuracy.  Even more accurate results are obtained computing,
for example by {\it simplified } \cite{BCsi} DAs, (biased with
$K_{1c}(R)$ {\it and} with the correction-to-scaling exponent
$\omega$), the ratio of the log-derivatives of the quantities ${\rm
m}_{\bot}^{(2)}(K_{1},R)/K_1$ and $\chi(K_1,R)$, which yields the
ratio $\nu_{\bot}(R)/\gamma(R)$. As it is also shown in
Fig.\ref{esprfm}, the estimates so obtained for this ratio appear to
be independent of $R$ to within $0.1\%$ along the whole interval
$0. \lesssim R \lesssim 1.5$, in which they remain quite near to the
ratio of the data\cite{xypisa} chosen for comparison. Of course, this
particularly favorable result is simply due to the fact that the
relative deviations of $\gamma(R)$ and $\nu_{\bot}(R)$, with respect
to their values at $R_M$, keep the same sign and a similar size as $R$
varies.

A blown-up view of part of these results is presented in
Fig.\ref{esponormal}, where we have plotted our temperature-biased DA
estimates of the exponents $\gamma(R)$, $\nu_{\bot}(R)$ and of their
ratio, after normalizing them to the corresponding
comparison\cite{xypisa} values.

It is amusing to remark that, for positive and not too large values of
$R$, our rough determination of the behavior of $a_{\chi} (R)$ also
suggests a simple prescription to improve the estimates of the exponents by
temperature-biased DAs.  We have simply  to correct for the expected
errors in the bias values of $K_{1c}(R)$, computing the critical
exponents by DAs biased with $K_{1c}(R) + \delta K_{1c}(R)/2$ when
$R>R_M+0.03$, or with $K_{1c}(R) - \delta K_{1c}(R)/2$ when $R<R_M-0.03$. We
have shown in Fig.\ref{espuniv} that this quite naive adjustment of
the standard biasing procedure to account for the sign and size of
$a_{\chi} (R)$ improves visibly the universality of the exponents with
respect to $R$.  At the same time, this result gives further support
to our conjecture that the spread $\delta K_{1c}(R)$ of the DA
estimates of $K_{1c}(R)$ is a sound approximation of their
uncertainty, provided that $R$ is positive and not too large.

\begin{figure}[tbp]
\begin{center}
\leavevmode
\includegraphics[width=3.37 in]{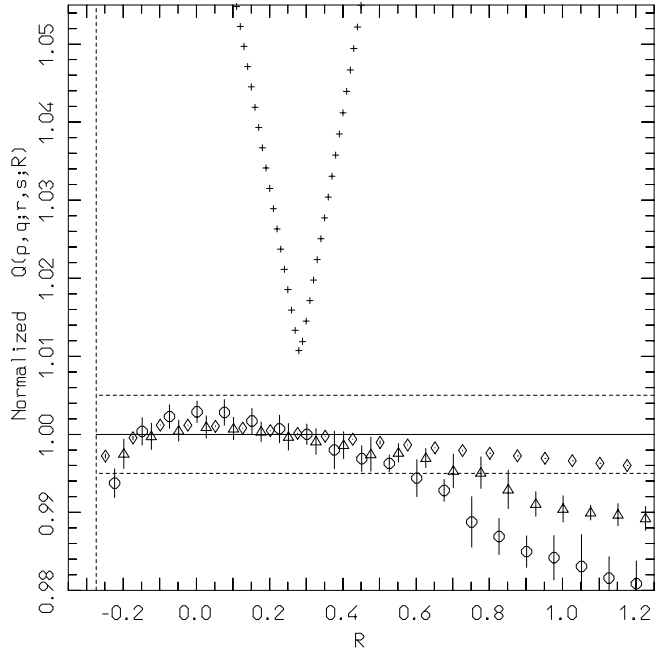}
\caption{ \label{figq} Universality with respect to $R$ of the moment
 ratios $Q(p,q;r,s;R)$.  We have reported results for $Q(1/2,1/2;0,1;R)$
 represented by circles, for $Q(1/2,1/4;0,3/4;R)$
 (triangles) and for $Q(1/2,1/2;1/4,3/4;R)$ (rhombs).
 The ratios are normalized to their values at $R_M$.  The values of
 these ratios at $R_M$ are respectively: $Q(1/2,1/2;0,1;R_M)=0.876(1)$,
 $Q(1/2,1/4;0,3/4;R_M)=0.929(1) $  and $Q(1/2,1/2;1/4,3/4;R_m)=0.9695(1) $. 
The sequence of crosses represents a quantity proportional to the the
 absolute value $|a^+_{\chi}(R)| $ of the correction-to-scaling
 amplitude (and is shifted upwards by 1.01 for graphical convenience).
The vertical dashed line indicates the value of $R_{LP}$.}
\end{center}
\end{figure}

 We have also studied other indicators of universality, such as, for
example, the values at the critical point $K_{1c}(R)$ of the
correlation-moment ratios
\begin{equation}
Q(p,q;r,s;R)=\frac{{\rm m}^{(p)}(K_{1},R){\rm m}^{(q)}(K_{1},R)} 
 {{\rm m}^{(r)}(K_{1},R){\rm m}^{(s)}(K_{1},R)}
\label{momrat}
\end{equation}
  with $p+q=r+s$. As expected, they show an approximate independence on
    $R$ for $R>0$.  This is shown in Fig.\ref{figq}, where we have
    plotted vs $R$ a few  ratios $Q(p,q;r,s;R)$ evaluated at the
    critical point and normalized to their value $Q(p,q;r,s;R_M)$ at
    $R_M$.   Our estimates refer to  the cases in which
    $(q=p=1/2; r=1/4,s=3/4)$, or $(q=1/2,p=1/4; r=0,s=3/4)$ or
    $(q=p=1/2; r=0,s=1)$.

  As long as $R$ is
positive and not too large, we can conclude that all these results
consistently and rather convincingly indicate that, along the {\it P-F} branch of the critical line, the small
violations of the exponent universality with respect to $R$, shown by
our numerical computations, are only apparent and can be entirely
ascribed to the slow convergence of approximation procedures still
unable, at the present orders of expansion, to account fully for the
presence of corrections to scaling.
 
   These evidences of universality with respect to $R$ justify the
technique\cite{Zinn81,Fisher} that we can adopt in order to improve
the accuracy in the determination of the exponents for the XY
universality class. We can observe that, as $R$ varies in the
ferromagnetic range, we have an $R$-dependent family of models all of
which can be assumed to belong to the XY universality class, so that
they share the same critical exponents, while they have generally
different $R$-dependent (namely non-universal) amplitudes of the
corrections to scaling.  Therefore, we expect that the best
approximations for the universal quantities will be achieved from the
study of the model with $R=R_M$, because $a^+_{\chi}(R_M)$ vanishes.
For this particular model in the family also the other leading
correction amplitudes of interest, for example $a^+_{\xi^2}(R)$, must
vanish at $R_M$, since the correction-amplitude ratios such as
$a^+_{\chi}(R)/a^+_{\xi^2}(R)$ are universal.

   These arguments support our belief that our exponent estimates
   $\gamma(R_M)=1.3177(5)$, $\nu(R_M)=0.6726(8)$ and
   $\nu(R_M)/\gamma(R_M)=0.5100(1)$ should be rated as the best
   possible determinations of the susceptibility and
   correlation-length exponents in the XY universality class, that one
   can extract, at the present expansion order, from our $R$-dependent
   family of HT series.

  The high-precision estimates\cite{xypisa} of the XY universality
   class exponents that we have compared to our results in
   Fig.\ref{esprfm}, were also obtained using a similar improvement
   procedure in the case of a different family of $nn$-interaction
   models for which series of order 23 are known.  Our best estimate
   of the susceptibility exponent agrees well with the corresponding
   result cited for comparison, although the uncertainty of our result
  is sizably
   larger, due to the still moderate length of our bivariate series.
   On the other hand, the central value of our best estimate for the
   exponent $\nu$ is somewhat larger than (while the estimate
   $\nu=0.6720(4) $ obtained from those of $\gamma$ and  of the ratio 
   $\nu/\gamma$
   is much closer to) the corresponding comparison value. Our direct
   estimate of $\nu$ is closer to the older MC result\cite{Hase}
   $\nu=0.6723(3)[8]$ and to the result $\nu=0.6717(3)$ of the
   simulation of Refs.[\onlinecite{buro,beach}].  It is, however, much
   larger than the value $\nu=0.6709(1)$ obtained, by using the
   hyperscaling relation $2-\alpha= d\nu$, from the recent
   high-accuracy measure $\alpha= 0.0127(3)$ of the $^4\rm He$
   specific heat in a micro-gravity experiment\cite{Lipa}.

  In Ref.[\onlinecite{buro}]
   recent determinations of $\nu$ with increasing accuracy have been
   summarized into a useful diagram showing that the central estimates
   from the most recent MC simulations and HT series analyses are, in
   general, significantly larger than those obtained both from the
   renormalization group and from the cited experimental measure. This
   is an interesting remark which calls, at least, for a more accurate
   reassessment of the uncertainties of the results in the literature.
   As far as our HT approach is concerned, we can reasonably expect
   that an extension of the bivariate expansions by only a couple of
   orders would significantly reduce the uncertainties of our best
   estimates of the exponents, particularly so for the direct estimate
   of $\nu$.  Presently, however, it might be less difficult to give
   further support to our arguments and make them sharper by using our
   series results as a guide for a high-precision MC simulation of the
   ANNNXY model  at  $R=R_M$.

   Finally we must add that, unfortunately, our series for the
 specific heat, which is a very weakly singular quantity, seems to be
 still insufficiently long to yield an evaluation of comparable
 accuracy for the exponent $\alpha(R)$.  We can only infer that on the
 {\it P-F} line, $\alpha(R) \simeq -0.01(2)$, as expected.  This
 result is completely compatible with the bounds 
$ -0.0202 <\alpha< -0.0124 $ 
obtained introducing into the hyperscaling relation
 the extremal values of the  range of 
 estimates of $\nu$ reported in the recent literature.

  We can conclude our study of the critical behavior along the {\it
 P-F} line remarking that from a study of $T_c(R)$ in a vicinity of
 $T_{LP}$, we can estimate $\phi=1.00(4)$, since it appears that the
 curve $T_c=T_c(R)$ changes from concave to convex in a small vicinity
 of $T(R_{LP})$.

\subsection{ The Lifshitz point and the {\it P-M} branch of the critical line}

 In the mean-field approximation $R_{LP}=-1/4$, but our HT calculation
shifts this value by $\simeq 10\%$ to the modulated side of the phase
diagram.  For $R_{LP} < R \lesssim 0 $, the critical exponents
crossover\cite{nasser} from the value of the XY universality class to
the LP critical behavior.  In this range of $R$, the amplitudes
$a^+_{\chi}(R)$ and $a^+_{\xi^2}(R)$ appearing in eqs.(\ref{chias})
 and (\ref{xi2as}) lose their leading role,
because also higher order correction amplitudes become important, the
convergence of our series slows down and they appear inadequate to
exhibit the universality with respect to $R$ of the exponents and
their expected discontinuous change to the values of the LP
universality class at $R=R_{LP}$. Thus, of course, in the crossover
region, also the spreads of our estimates will grossly underestimate
the real errors.

\begin{figure}[tbp]
\begin{center}
\leavevmode
\includegraphics[width=3.37 in]{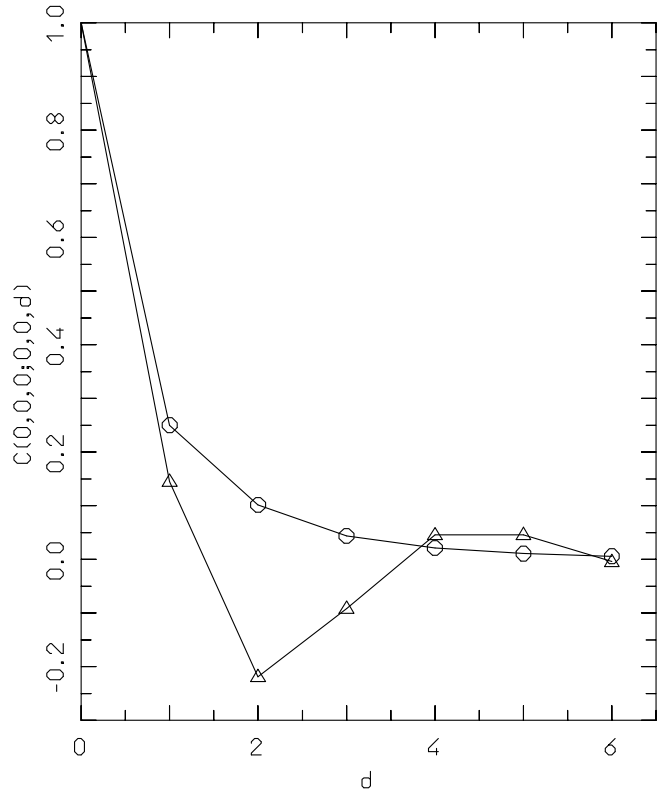}
\caption{ \label{corlindis} The correlation function $C(0,0,0;0,0,d)$
 between the spin at the origin and a spin on the $z$-axis as a
 function of the distance $d$ between the spins, at fixed values of
 $R$ and $T$.  We have chosen the value $(R=0.11, T_c(R)/2=2.59)$
 (open circles) 
 on the right-hand of the disorder line and the value 
 $(R=-1.0,T_c(R)/2 =2.439)$(open triangles) 
 on the left-hand of it, to display the different behavior of the
 corresponding correlation function (pure exponential decay  and
 oscillating exponential decay respectively). 
The lines connecting the symbols are drawn as a guide for the eye.}
\end{center}
\end{figure}

\begin{figure}[htbp]
\unitlength=1in
\begin{center}
\includegraphics[width=3.37 in]{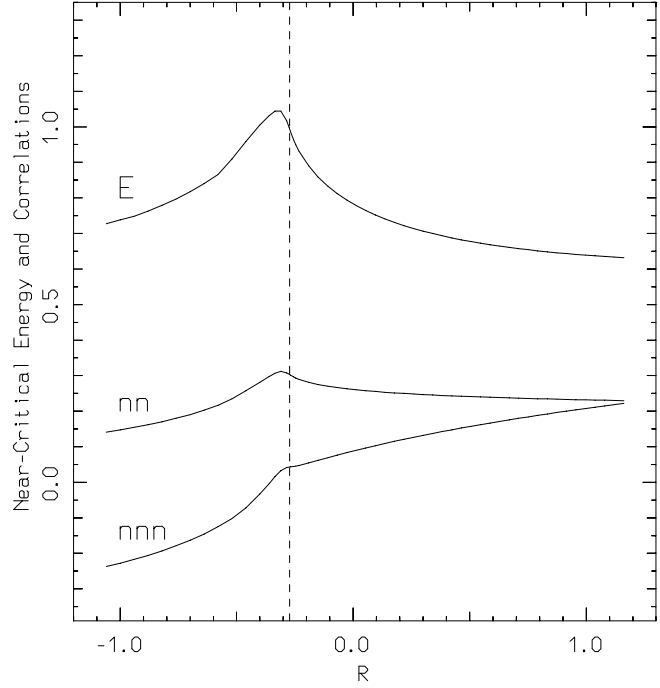}
\caption{ \label{energ} The energy (upper curve), the $nn$-spin
 correlation function
 (middle curve) and the $nnn$-spin correlation function (lowest curve)
 computed near the boundary of the paramagnetic phase (at $T= 1.1 T_c(R)$)
 and plotted vs $R$. The vertical dashed line indicates the value 
of $R_{LP}$. }
\end{center}
\end{figure}

Let us now recall that the eq. ${\rm m}^{(2)}_{\|}(K_1,R)=0$
implicitly defines in the $R-T$ plane the so-called {\it disorder}
line\cite{Step}, which, within the paramagnetic phase, divides a
region with monotonically (exponentially) decaying correlations along
the direction parallel to the $nnn$ interaction from a region with
oscillating, but still exponentially damped correlations.  The
equation defining the disorder line can be solved iteratively with
respect to $R$ to form the single-variable series
$R_{dis}=R_{dis}(T)$, which finally is re-summed by DAs. The plot of
the disorder line obtained from this series is drawn as an almost
vertical dashed line in Figure \ref{bcrfm}.  It may be of interest to
show how accurately the spin-spin correlations can be computed at HT
and therefore we have displayed in Fig.\ref{corlindis} the qualitative
difference in their behavior as functions of the distance of the spins
along the $z$-direction, on the two sides of the disorder line.  Using
our knowledge of the spin-spin correlations, we can also show that the
LT modulated order already begins to build up in the nearly critical
HT phase. This is suggested by Fig.\ref{energ}, where we have plotted vs
$R$ the values of the energy, the $nn$ and the $nnn$ spin correlations
along the {\it z-}axis, calculated just above the boundary of the
paramagnetic phase, precisely at $T= 1.1 T_c(R)$.  We can observe that
the energy reaches a maximum near $R_{LP}$, where the disorder is
higher and that the $nn$ spins are positively correlated on the whole
range of $R$ (albeit not too strongly since $T$ is high). On the other
hand the $nnn$ spins tend to be more correlated than the $nn$ spins
for $R \gtrsim 1.2$, while as $R$ is lowered, this correlation decays to
become negative when $R \lesssim R_{LP}$.
 
 In order to locate the LP on the boundary of the paramagnetic phase,
 we have to recall\cite{Redner,Ferer} that for small $q_z$
 \begin{equation}
  \chi(0,K_1,R)/\chi({ q_z},K_1,R)=  1+
q_z^2 \frac {  {\rm  m}^{(2)}_{\|}(K_1,R)} {2 \chi(K_1,R)} +
 q_z^4 \big(\frac { {\rm  m}^{(2)}_{\|}(K_1,R)^2}{4 \chi(K_1,R)^2} -
 \frac { { \rm  m}^{(4)}_{\|}(K_1,R)} {24 \chi(K_1,R)}\big)+.. 
\label{qsmall}
\end{equation}
showing that the minimum at $q_z=0$, which characterizes
$\chi(0,K_1,R)/\chi({ q_z},K_1,R)$ at fixed $K_1$, when $R$ is in the
ferromagnetic range, changes to a local maximum as $R \rightarrow
R_{dis}(K_1) $ where the second-order parallel moment ${\rm
m}^{(2)}_{\|}(K_1,R)=0$.  Thus the LP is found at the intersection of
the {\it P-F} branch of the critical line with the disorder
line. Following this procedure we are led to the estimate
$(R_{LP}=-0.2733(6) $, $T(R_{LP})=1.778(2))$ of the intersection point
between the critical locus and the disorder line.  The value of
$R_{LP}$ obtained in this way is consistent with that obtained
minimizing $\chi(0,K_1,R)/\chi({ q_z},K_1,R)$ with respect to $q_z^2$
as $R-R_{LP} \rightarrow 0^-$ and $K_1 \rightarrow K_1(R)$.  Indeed,
for small $q_z^2$, we obtain from eq.(\ref{qsmall}) that the position
of the minimum $\bar q_{z}^2$ vs $R$ is given by
 \begin{equation}
  \bar q_{z}^2 \approx \frac {6{\rm  m}^{(2)}_{\|}(K_1,R) \chi(K_1,R)}
 { { \rm  m}^{(4)}_{\|}(K_1,R)\chi(K_1,R) -6{\rm  m}^{(2)}_{\|}(K_1,R)^2 } 
\label{q2small}
\end{equation}
 evaluated at the critical point $K_{1c}=K_{1c}(R)$ for $R \lesssim R_{LP} $.
 By re-summing the series expansion of $\bar q_{z}^2$, 
 we can determine  $R_{LP}$, also as
 the value of $R$ at which $\bar q_{z}^2=0$.

\begin{figure}[htbp]
\unitlength=1in
\begin{center}
\includegraphics[width=3.37 in]{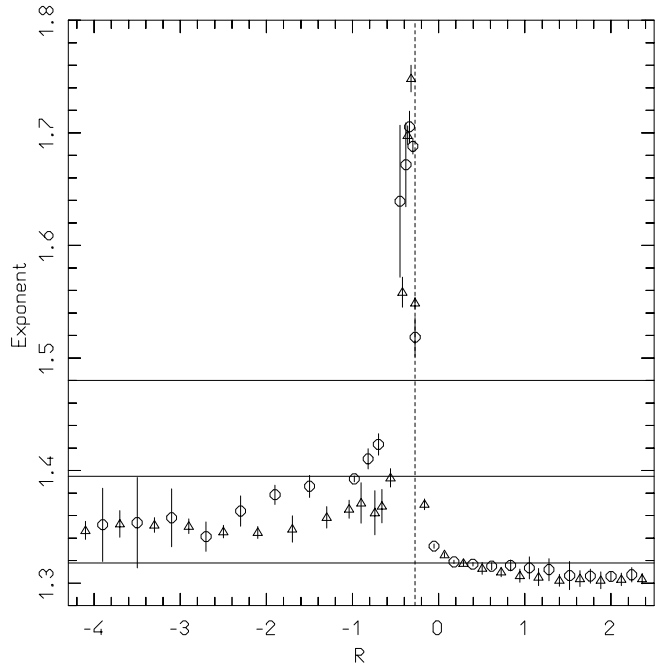}
\caption{ \label{ga3} The exponent $\gamma(R)$ computed along 
the {\it P-M} branch of the critical line. The estimates obtained by
  the critical-point renormalization method  are represented as open triangles,
 those obtained by DAs  biased with the critical temperature as open circles.
The horizontal continuous lines represent the values of 
the exponent $\gamma$ for
 the  $O(2)$ (lowest line), $O(3)$ (intermediate line) and  $O(4)$
 (upper line) universality classes.
The vertical dashed line indicates the value of $R_{LP}$. }
\end{center}
\end{figure}

 For $R \lesssim 0$, in the crossover region, Figs.\ref{esprfm} and \ref{ga3}
 show a steep rise of our exponent estimates near the LP and past it,
 followed by a somewhat slower decay extending through $R \simeq-0.8$.
 A completely similar behavior of the exponent $\gamma(R)$ nearby the
 LP was noted in a HT study\cite{Ferer} of the 3d ANNNI model. In this
 study, we should then assume that the values of the exponents at the
 LP are not strongly affected by the crossover and also be aware that
 they are rather sensitive to the location of the LP.  Using
 the critical-point renormalization method, we can estimate
 $\gamma_l=1.55(1)$, while the biased DAs suggest $\gamma_l=1.52(1)$,
 as shown in Fig.\ref{esprfm}.  We shall take a weighted average of
 these values as our final estimate of the exponent:
 $\gamma_l=1.535\pm 0.025 \pm0.2|R_{LP}+0.2733|$,
 including explicitly in our error a contribution from 
the uncertainty of $R_{LP}$.  This result has a smaller uncertainty than but
 is completely compatible with the estimate $\gamma_l=1.5(1)$ of the
 MC simulation of Ref.[\onlinecite{Selke80}]. Notice that our
 estimates of $R_{LP}$ and $T(R_{LP})/2.$ differ by $\simeq 5\%$ from
 the values $R_{LP}= -0.263(2)$ and $T_{LP}(R)/2.  \simeq 1.82$,
 determined by the old (sixth-order) series\cite{Redner} and used as
 an input in the MC\cite{Selke78,Selke80} study.  Our estimate of the
 exponent is also not far from the value $\gamma_l=1.495 $
 obtained\cite{DiSh,Shpot} simply by setting $\epsilon= d_u(1)-d=3/2$
 in the two-loop $\epsilon$-expansion.

  Starting with the HT expansions of $ \xi^2_{\bot}(K_1,R)$, similar
 considerations yield the estimate 
 $\nu_{\bot}=0.805\pm 0.015 \pm 0.1|R_{LP}+0.2733|$.  For this
 exponent, no MC results are available and our result can be compared
 only with the value $\nu_{\bot}=0.757$ from the
 $\epsilon$-expansion. 

  The direct estimate of $\nu_{\|}$ from the analysis of $\xi_{\|}$ is
   notoriously difficult because, in the {\it P} phase near the
   Lifshitz point, the competition between the $nn$ and the $nnn$
   interaction reduces drastically the correlation length in the $z$
   direction. In particular, in the $N=1$ case, the determination of
   $\nu_{\|}$ has so far eluded even the most extensive\cite{PHENK} MC
   simulation so far available. 
 In the large $N$ limit (namely in the case of the
   uniaxial {\it spherical} model), in which very long HT expansions
   can be easily computed, it was observed\cite{Redner} that at least
   35 orders are necessary to approximate the behavior of $\xi_{\|}$!
   Also in the $N=2$ case under study, the length of our HT
   expansions, unfortunately, seems to be still insufficient.  We can,
   however, try to estimate indirectly $\nu_{\|}$, either from a
   measure of the exponents $\beta_q$ and $\phi$, taking advantage of
   the scaling law $\nu_{\|}=\beta_q \phi$, or by determining the
   exponent $\alpha_l$ and then using the generalized hyperscaling law
   along with our previous estimate of $\nu_{\bot}$.  As above
   remarked, the determination of the exponent $\alpha_l$ from our
   series is not yet accurate enough to be  useful. It does
   suggest, however, that $\alpha_l=-0.02(2)$, which is compatible
   with the $\epsilon$-expansion estimate in Table \ref{tab1}.  If,
   quite conservatively, we simply assume that $\alpha_l$ is negative
   and $|\alpha_l| < 0.1$, we get from the hyperscaling law the
   rough bounds $ 0.36 <\nu_{\|}<0.52$, which are consistent with the
   $O(\epsilon^2)$ value $\nu_{\|} \simeq 0.372 $ reported in table
   \ref{tab1}.  On the other hand, using the above reported estimate
   $\phi= 1.00(4)$ and the estimate  $\beta_q= 0.40(2)$ given below, we
   get the value $\nu_{\|}=0.40(3)$.  Plugging back this result into
   the hyperscaling law, we find the value $\alpha_l = -0.01(6)$ for
   the specific-heat exponent, which is also compatible with all above
   indicated estimates. The values of the remaining exponents can all be
   obtained from the scaling laws.

  We can thus conclude that our HT determinations of the critical
 exponents at the LP are in most cases consistent, to within 
 $\simeq 10\%$, with the $\epsilon-$expansion estimates.

In order to map out the branch of the critical line with $R < R_{LP}$,
 we have to analyze the wave-vector-dependent susceptibility
 $\chi({q_z},K_1,R)$.  For $R<R_{LP}$ the critical point $K_{1c}(R)$
 can be determined by locating the nearest positive singularity in
 $K_1$ of $ \chi({q_z},K_1,R)$ with $q_z$ near the peak value
 $\bar q_z=\bar q_z(K_1,R)$.  The value of $K_{1c}(R)$ is taken to be
 the minimum with respect to $q_z$ of the singularity locus
 $K_1=K_1(q_z,R)$ of $\chi({q_z},K_1,R)$.  Our results for the {\it
 P-M} branch of the critical line, also drawn in Fig.\ref{bcrfm},
 complete the map of the boundary of the paramagnetic phase. In the
 same figure, we have schematically indicated a transition line, 
 which separates the ferromagnetic and the modulated phases
 and joins the LP to the point $R=-1/4,T=0$, where the ordering of the
 ground state changes from ferromagnetic to modulated. This line is also
 expected to be of second order, but is beyond the reach of our HT
 analysis.  Our phase
 diagram agrees well with the cited results\cite{Selke78} obtained
 supplementing a MC simulation with the sixth-order HT expansions of
 Ref.[\onlinecite{Redner}].

  It must be noted that the uncertainties of the points of the {\it
  P-M} branch of the critical line are sizably larger than those of
  the {\it P-F} branch, making it more difficult to obtain precise
  temperature-biased estimates of the critical exponents. If,
  nevertheless, we insist in computing some rough estimate of the
  exponent $\gamma(R)$, it is encouraging to observe that the results
  obtained from temperature-biased DAs remain essentially consistent,
  over a wide range of values of $R$, with those computed by the
  critical-point renormalization method which is insensitive to the
  uncertainties of the critical temperatures. The results for
  $\gamma(R)$ obtained by these two methods are reported in
  Fig.\ref{ga3}.  For $R< -0.8$, on the left of the region of
  crossover from the value of $\gamma_l$, we observe that both
  sequences of estimates tend to stabilize at some value which is
  intermediate between those of the $O(2)$ and of the $O(3)$
  universality classes.   These puzzling results might deserve
  further confirmation by a MC study. If they are confirmed, the MC
  approach\cite{Diep2} would also be best suited to investigate
  whether a possible (weak) first-order\cite{Tiss} (rather than
  second-order) character of the {\it P-M} transition might explain
  its features. This possibility was suggested by a renormalization
  group study of a Landau-Ginzburg effective four-component model of a
  biaxial $N=2$ system\cite{Garel} which turned out to have only
  unstable fixed points at second order in the $\epsilon$-expansion.

\begin{figure}[htbp]
\unitlength=1in
\begin{center}
\includegraphics[width=3.37 in]{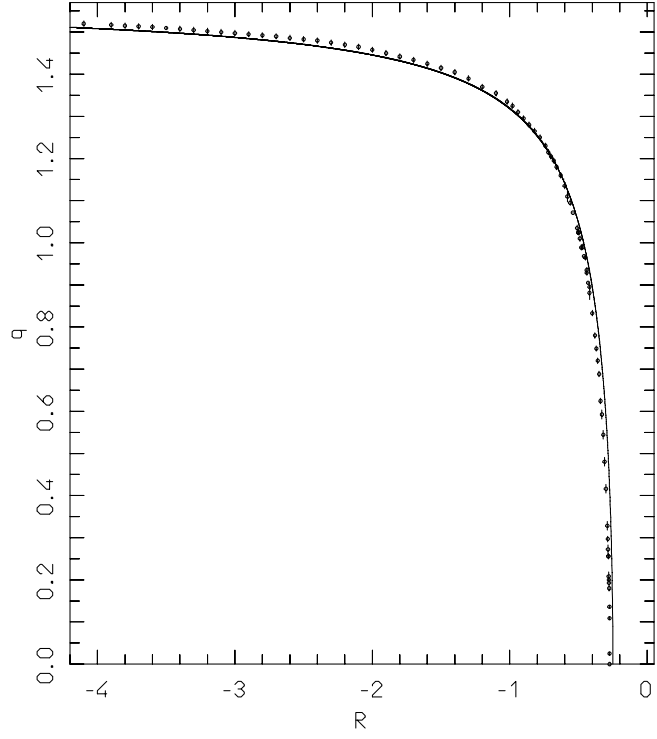}
\caption{ \label{qvsR} The peak value of the modulation vector $\bar q_z$ 
at the  critical temperature plotted vs. $R$ for $R \leq R_{LP}$. 
The continuous curve represents the
mean-field prediction $\bar q^{MF}_z= cos^{-1}(R^{MF}_{LP}/R)$ with 
$R^{MF}_{LP}=-1/4$. }
\end{center}
\end{figure}

The curve describing the peak value $\bar q_z$ of the modulation
wave-number $q_z$ at $T=T_c(R)$ vs $R$, obtained from our analysis, is
reported in Fig.\ref{qvsR} and compared to the mean-field prediction
$\bar q_z^{MF}(R)=cos^{-1}(R^{MF}_{LP}/R)$ with $R^{MF}_{LP}=-1/4$.
As pointed out in Ref.[\onlinecite{Redner}], also in the Ising case,
at high temperature the peak of $\chi({q_z},K_1,R)$ occurs at 
$\bar q_z^{MF}(R)$ but, as the temperature decreases to $T_c(R)$, the peak
moves to lower values of $q_z$ for $R \gtrsim -0.6$ or otherwise to
slightly higher values. This is clearly shown in Fig.\ref{qvsR}.  The
largest deviations of our results from the mean-value approximation
occur in a small vicinity of $ R_{LP}$.

From the behavior of this curve near $R \simeq R_{LP}$, we have  estimated
the value $\beta_q= 0.40(2)$ used above.

\section{Acknowledgements}
This work was partially supported  by the MIUR.  We also thank
  the Physics Depts. of Milano-Bicocca University and of Milano
  University for their hospitality and support.  We are very grateful
  to Prof. W. Selke for critically reading the first draft of this
  paper and for calling our attention to some references.  Our
  computations have been mostly performed by the {\it  Turing} pc-cluster 
  of the Milano-Bicocca INFN Section. We also thank
  Prof. L.N. Shchur for granting us access to the {\it PARMA} cluster  of the
  Landau Institute in Chernogolovka and to Dr. Maria Medvedeva for her
  assistance.

\appendix

\section{ Derivation and validation of the series}

 We have used the algorithm\cite{SDEM} of the Schwinger-Dyson
   equations(SDE) to compute the HT expansions for the $N$-vector spin
   models under study.  When it was introduced, the SDE method could
   be profitably applied only to systems with $nn$
   interactions\cite{SDEA}, due to the severe limitations in the
   memory and speed of the computers available two decades ago.  This
   method was repeatedly described in Refs.[\onlinecite{SDEA,SDED}],
   and therefore it is sufficient to recall only that for the XY
   models described by the Hamiltonians (\ref{tnhamilt}) and
   (\ref{snhamilt}), the SDE take a particularly simple and suggestive
   form.  Indicating by 
 ${\bf s}= ({\vec x_1},q_1;{\vec  x_2},q_2;...{\vec x_k},q_k)$ 
  a set of site coordinates $\vec x_i$
   and of integer nonzero quantities $q_i$ attached to them and such
   that $\sum^k_{i=1} q_i=0$, the generic correlation function can be
   written as
\begin{equation}
<\phi({\bf s})> = \frac{1} {Z} \int \Pi_x d\theta_x \phi({\bf s}) e^{-\beta H}
\label{gfcor}
\end{equation}
 where $\phi({\bf s})= \exp[i\sum_k q_k \theta_{x_k}]$ and $\theta_x$
is the angle formed by the unit vector ${\vec v(\vec x)}$ with a fixed
direction.  The SDE are:
\begin{equation}
<\phi({\bf s})> = -\frac{K_1} {q_1} \sum_{\mu} (<\phi({\bf s^+_{\mu}})> -
<\phi({\bf s^-_{\mu}})>)  -\frac{K_2} {q_1} \sum_{\nu} 
(<\phi({\bf s^+_{\nu}})> -<\phi({\bf s^-_{\nu}})>)
\label{sdefcor}
\end{equation}
Here 
\begin{equation}
\phi({\bf s^+_{\mu}})= \exp(i\theta_{x_1})
\exp(-i\theta_{x_1+a_{\mu}}) \phi({\bf s}),
\end{equation} 
while 
\begin{equation}
\phi({\bf
s^-_{\mu}})=\exp(-i\theta_{x_1}) \exp(i\theta_{x_1+a_{\mu}}) \phi({\bf
s}). 
\end{equation}
Moreover
\begin{equation}
\phi({\bf s^+_{\nu}})= \exp(i\theta_{x_1})
\exp(-i\theta_{x_1+b_{\nu}}) \phi({\bf s}) , 
\end{equation}
and
\begin{equation}
 \phi({\bf
s^-_{\nu}})=\exp(-i\theta_{x_1}) \exp(i\theta_{x_1+b_{\nu}}) \phi({\bf
s})  
\end{equation}
with ${\vec x_1}$ an arbitrary site in the set ${\bf s}$.  In the
first sum of eq.(\ref{sdefcor}), $a_{\mu}$ is the lattice vector
joining the site $\vec x_1$ to the $nn$ sites, whereas in the second
sum $b_{\nu}$ denotes the vector joining the site $\vec x_1$ to the
$nnn$ sites in the case of the Hamiltonian eq. (\ref{tnhamilt}) (the
vector joining the site $\vec x_1$ to the $sn$ sites in the case of
the Hamiltonian eq. (\ref{snhamilt})).

The computational complexity of the SDE method increases with the
lattice dimensionality and, for a given dimensionality, with the
effective coordination-number, i.e.  as the number of interacting
neighbors increases.  Essentially the same difficulty is met with the
evaluation of the embedding factors in the conventional graph
approach, but our method has the advantage of using the SDE as
recurrence relations among the correlation functions and of requiring
only a straightforward iteration of these relations, which avoids
altogether all combinatorial intricacies of the graph method.  In the
specific cases under study, due to the presence of the $nnn$
interactions, this method is, by some orders of magnitude, more memory
and CPU demanding than for the pure $nn$ interactions, but otherwise
not much more difficult.  Thus for example, in the computationally
most intensive case: that of the Hamiltonian of eq.(\ref{tnhamilt}) in
$3d$ with $3-axial$ interaction, the general SDE is a linear
relationship among 25 {\it a priori} different, correlation functions,
while in the simpler $nn$ interaction case in three-dimensions, the
SDE involve only 13 correlation functions.

A carefully designed code for the Schwinger-Dyson equations can
compute moderately long series using only a reasonable CPU time of an
ordinary 32-bit single-core processor desktop PC.  In particular, our
codes can reproduce all previously existing series data in a
negligible CPU time of the order of $10^{-3}$ sec.  Due to the
sensitive dependence of the computational load on the effective
coordination number, the series that we could derive for the $1-axial$
models are longer (and more fastly computed) than those for the
$d-axial$ models.  In the case of the uniaxial model studied here, the
expansion through order 17 was completed in approximately three weeks
by a PC, while the order 18 was obtained by using a few nodes of a PC
cluster for an equivalent single-processor CPU-time of approximately
five months.  Longer series might be obtained by a more extensive
parallelization of our codes or, perhaps, by returning to the
conventional graph methods, provided that the efficiency of the
current graph-embedding algorithms can be drastically improved.
 
  The comparison of the extended expansions with independent,
   sufficiently long and reliable previous results is always a
   necessary step in the validation of the codes and of the results
   for automated derivations.  Generally, in our case, this was not
   satisfactorily feasible.  In two dimensions, a comparison was
   possible only for the {\it 2-axial} model (\ref{tnhamilt}), for
   which a fifth-order expansion of the susceptibility was tabulated
   in Ref.[\onlinecite{Clarizia}] and in three-dimensions, only in the
   case of the {\it 1-axial} model, for which an expansion of the
   susceptibility was tabulated through sixth order in
   Ref.[\onlinecite{Redner}].  Our series agree with these 
   results.  Due to the lack of other published data and to the
   present unavailability of old short unpublished series
   \cite{Rednerpc} for the {\it 3-axial} case in three dimensions, no
   further comparison with independent calculations was possible.
   Weaker (and obvious) tests, through all orders that we have computed, are
   feasible in the limiting cases\cite {bc} in which one of the
   exchange interaction constants vanishes.  For example, when
   $K_2=0$, the series for the {\it m-axial} models should reduce to
   those of the $nn$ interaction models on the same lattice.  Therefore the 
   series for the {\it 1-axial} models should reduce to those of a
   $nn$ interaction model on the same lattice when $K_2=0$, and to
   those of a $nn$ model on a $1d$ lattice when $K_1=0$. Of course,
   our results pass also these tests, which however pin down only 2
   out of the $l+1$ coefficients of each order $l$.  Our confidence in
   the correctness of the calculations, for which we have written two
   sets of largely independent codes, in Fortran and in C++, receives
   further support also from the stability of the numerical results
   under our many rewritings of both sets of codes to increase their
   efficiency, as well as from the smooth and consistent behavior of
   the quantities analyzed.

\newpage

\begin{figure}[tbp]
\begin{center}
\leavevmode
\includegraphics[width=3.37 in]{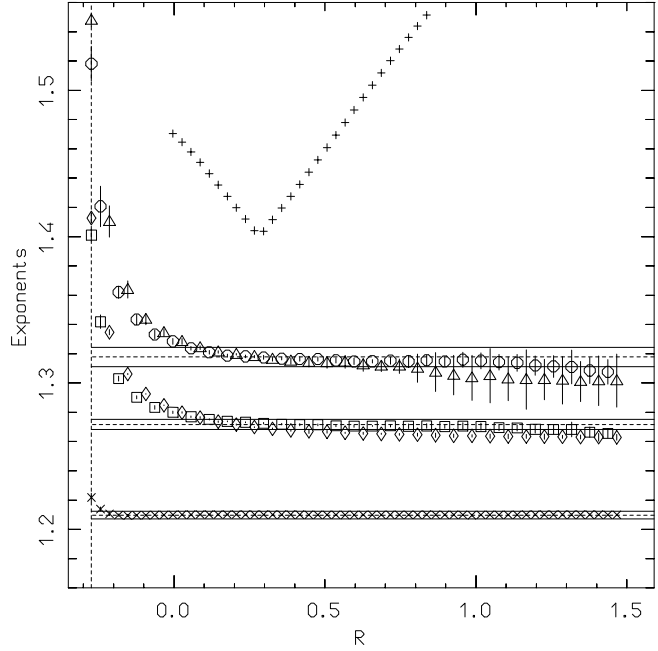}
\caption{ \label{esprfm} The exponents $\gamma(R)$ and $\nu_{\bot}(R)$
 (notice that for $R \gtrsim 0$ $\nu_{\bot}=\nu$) are computed along
 the {\it P-F} branch of the critical line $K_{1c}=K_{1c}(R)$ in order
 to display their universality with respect to $R$.  The horizontal
 continuous lines are bands of 0.5 $\%$ deviation from the central
 values, indicated by dashed lines, of recent high-precision
 estimates\cite{xypisa} of the exponents $\gamma$ (in the case of the
 upper band), $\nu$ (central band, shifted upwards by 0.6) and of the
 ratio $\nu/\gamma$ (lower band shifted upwards by 0.7) for the XY
 universality class.  We have indicated by open circles our estimates
 of the exponent $\gamma(R)$ obtained from DAs biased with the
 critical temperature, while the estimates obtained by the
 critical-point renormalization method, generally subject to a larger
 uncertainty, are denoted by open triangles.  In the case of the
 exponent $\nu_{\bot}(R)$ we have denoted by open rhombs the estimates
 obtained from DAs biased with the critical temperature and by open
 squares the estimates obtained by the critical-point renormalization
 method.  As already noticed, for graphical convenience, the values of
 this exponent are shifted upwards by 0.6.  Finally a sequence of
 stars denotes the ratios $\nu_{\bot}(R)/\gamma(R)$ which, again for
 convenience, are shifted upwards by 0.7.  A sequence of crosses
 schematically indicates a quantity proportional to the the absolute
 value $|a^+_{\chi}(R)| $ of the correction-to-scaling amplitude. For
 graphical convenience, this quantity is shifted upwards by 1.4.  The
 vertical dashed line on the left-hand side indicates the value
 $R_{LP}=-0.2733(5)$ corresponding to the LP.}
\end{center}
\end{figure}

\begin{figure}[tbp]
\begin{center}
\leavevmode
\includegraphics[width=3.37 in]{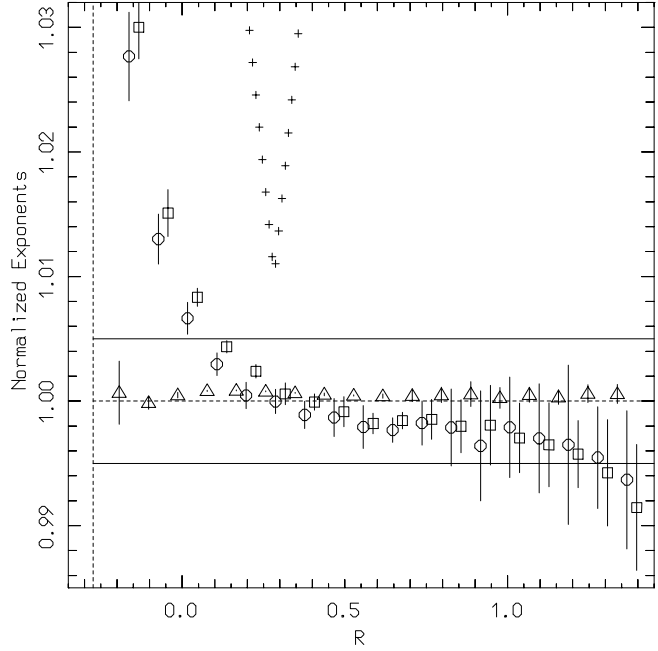}
\caption{ \label{esponormal} A slightly modified and blown up
representation of some of the data appearing in Fig.\ref{esprfm}. We report
the exponents $\gamma(R)$(open circles), $\nu_{\bot}(R)$(open squares)
and the ratio $\nu_{\bot}(R)/\gamma(R)$(open triangles) computed by
temperature-biased DAs along the {\it P-F} branch of the critical line
$K_{1c}=K_{1c}(R)$.  The data are now normalized to the central values
of the corresponding comparison estimates\cite{xypisa} of the
exponents for the XY universality class. The horizontal solid lines
are bands of 0.5 $\%$ deviation from the comparison value (central
dashed line).  The vertical dashed line on the left-hand side
indicates the value of $R_{LP}$. The upper curve, schematically denoted by
crosses, plots a quantity proportional to the the absolute value
$|a^+_{\chi}(R)| $ of the correction-to-scaling amplitude. This
quantity is shifted upwards by 1.01 for graphical convenience.}
\end{center}
\end{figure}

\begin{figure}[tbp]
\begin{center}
\leavevmode
\includegraphics[width=3.37 in]{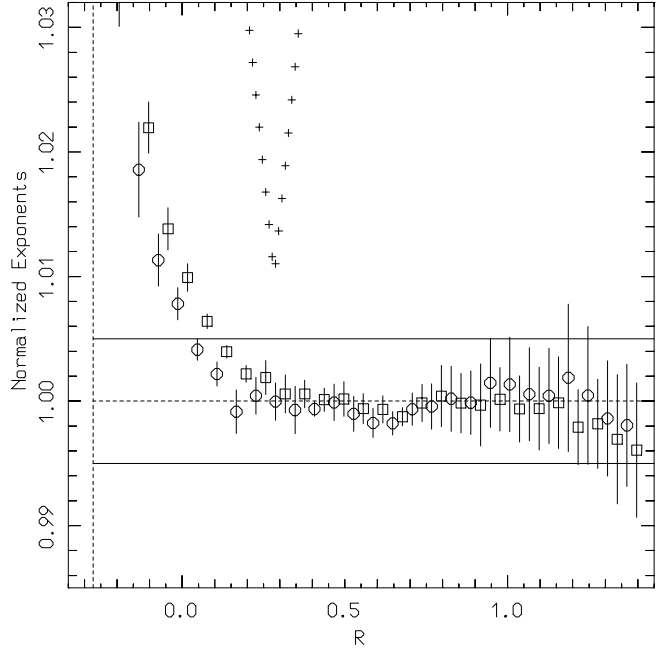}
\caption{ \label{espuniv} Same as Fig.\ref{esponormal}, but here the
 exponents $\gamma(R)$ (open circles) and $\nu_{\bot}(R)$ (open squares) are
 computed along the {\it P-F} branch of the critical line by DAs
 biased with $K_{1c}(R)+ \delta K_{1c}(R)/2$ for $R>R_M+0.03$ or biased
 with $K_{1c}(R)- \delta K_{1c}(R)/2$ for $R<R_M-0.03$. Like in
 Fig.\ref{esponormal}, the exponents are normalized to the
 mentioned\cite{xypisa} comparison central values $\gamma=1.3178$ and
 $\nu=0.6717$, respectively.  The horizontal solid lines are bands of
 0.5 $\%$ deviation from the central value indicated by a dashed line.
 The vertical dashed line indicates the value of $R_{LP}$.  The
 sequence of crosses represents a quantity proportional to the the
 absolute value $|a^+_{\chi}(R)| $ of the correction-to-scaling
 amplitude (shifted upwards by 1.01 for graphical convenience).  }
\end{center}
\end{figure}

\begin{table}
\caption{ Values of the critical exponents for the 
  $(m,d,N)=(1,3,2)$ (uniaxial XY) Lifshitz point in three dimensions.
  The column with heading MF, taken from Ref.[\onlinecite{DiSh}]
 reports the mean
  field exponents.  In the column with heading $O(\epsilon^2)$, 
 we report the values obtained\cite{DiSh} by
  evaluating at $\epsilon=3/2$ the $\epsilon-$expansions of the
  exponents truncated at the second order.  Under the heading HT we report
  the results directly obtained in this paper.  
  The heading MC refers to the results of
  the simulations in Refs.[\onlinecite{Selke78,Selke80}].}
\center
\begin{tabular}{|c|c|c|c|c|} \hline
Exponent  &  $MF $  &$O(\epsilon^2)$ & $HT$ & $MC$  \\\hline
$\gamma_l$&\phantom{++}1\phantom{++}&\phantom{++}1.495\phantom{++}
 &\phantom{++}1.535(25)\phantom{++}
 &\phantom{++}1.5(1)\phantom{++} \\
$\nu_{\|}$ &$\frac {1} {4}$&0.372 &0.40(3) & \\
$\nu_{\bot}$ &$\frac {1} {2}$&$0.757$&$0.805(15)$&  \\
$\eta_{\|}$ &0&-0.020&&  \\
$\eta_{\bot}$ &0&0.042&& \\
$\alpha_l$ &0&-0.047 & &$0.10(14)^*$\\
$\beta_l$ &$\frac {1} {2}$ &0.276 & &0.20(2)\\
$\delta_l$ &3 & & &\\
$\phi_l$ &$\frac {1} {2}$ &0.725& 1.00(4)&\\
$\beta_q$ &$\frac {1} {2}$&0.521& 0.40(2)&\\\hline
\end{tabular}
\label{tab1}
\end{table}

\begin{table}
\caption{ The critical values of $K_1$ for selected values of $R$.
 The uncertainties reported here correspond only to the spread of the
 DA estimates and therefore are likely to underestimate the real
 errors when $|R-R_M|$ is not small, particularly so for 
$R \lesssim 0$.}  \center
\begin{tabular}{|c|c|c|c|} \hline
 $R$  &  $K_{1c}(R)   $ & $R$  &  $K_{1c}(R)$  \\\hline
   1.200 &  0.15239(4) &   0.500 &  0.18429(1)\\
   1.100 &  0.15596(4) &   0.400 &  0.19075(1)\\
   1.00 &   0.15980(4) &  0.300 &  0.19799(1)\\
   0.900 &  0.16391(3) &     0.200 &  0.20623(1)\\
   0.800 &  0.16836(3) &     0.100 &  0.21577(1)\\
   0.700 &  0.17319(2) &     0.000 &  0.22710(2)\\
   0.600 &  0.17847(2) &    -0.100 &  0.24113(4)\\
\colrule  
\end{tabular}
\label{tabk}
\end{table}


\begin{thebibliography}{}

\bibitem[*]{pb} Electronic address: paolo.butera@mib.infn.it
\bibitem[**]{mp}Electronic address: mario.pernici@mi.infn.it

\bibitem{Dalton} N. W.~Dalton and D. W. ~Wood, J. Math. Phys. {\bf 10}, 
1271 (1969); G. ~Paul and H. E. ~Stanley, Phys. Lett. {\bf 37}A, 328 (1971); 
 G. ~Paul and H. E. ~Stanley, Phys. Rev. B {\bf 5}, 3715 (1972).

\bibitem{Philhours} J. Philhours, Phys. Rev. B {\bf 4}, 929 (1971).

\bibitem{Luijte} E. Luijten and H. W. J. ~Bl\"ote, 
Int. J. Mod. Phys. C {\bf 6}, 359 (1995).

\bibitem{Yosh}A.~Yoshimori, J. Phys. Soc. Japan {\bf 14}, 807 (1959); 
T. ~Kaplan, Phys. Rev. {\bf 116}, 888 (1959);
J.~Villain, J. Phys. Chem. Solids,  {\bf 11}, 303 (1959).

\bibitem{El} R. J. ~Elliott, Phys. Rev. {\bf 124}, 346 (1961).

\bibitem{SelFi} W. Selke and M.E. Fisher, Z. Phys.  B {\bf 40}, 71 (1980).

\bibitem{Step} J.~Stephenson and D. D. ~Betts, Phys. Rev. B {\bf 2}, 
2702 (1970); J.~Stephenson, Can. J. Phys. {\bf 48}, 1724 (1970); 
Phys. Rev. B {\bf 1}, 4405 (1970);  B {\bf 15}, 5442 (1977); 
  B {\bf 15}, 5453 (1977).

\bibitem{Hornreich} R.M. ~Hornreich, M. ~Luban and S. ~Shtrikmann,
 Phys. Rev. Lett. {\bf 35}, 1678 (1975); M. ~Droz and M.D. ~Coutinho-Filho,
 in {\it Proc. 21st Conf. on Magnetism and Magnetic Materials}, 
 Philadelphia (AIP, New York, 1975); L. ~Kalok and G.M. ~Obermair, J. Phys. C 
{\bf 9}, 818 (1976).  


\bibitem{Hor80} R.M. ~Hornreich, J. Magn. Magn. Mater.
 {\bf 15-18}, 387 (1980).  

\bibitem{Fith} M.E. ~Fisher, Physica A {\bf 106}, 28 (1981).

\bibitem{Lieb} R. ~Liebmann, 
{\it Statistical mechanics of periodic frustrated systems}, 
Lecture Notes in Physics Vol.251, 
 (Springer, Berlin 1986).

\bibitem{Selke} W. ~Selke, Phys. Rep. {\bf 170}, 213 (1988).

\bibitem{Seldl} W. ~Selke,
 in {\it Phase Transitions and critical Phenomena}, edited by  
 C.~Domb and J.~Lebowitz, (Academic, New York, 1992) Vol. 15.

\bibitem{Yeom} J. M.~Yeomans, Solid State Phys. {\bf 41}, 151(1988). 

\bibitem{Seu} M.~Seul and D.~Andelman, Science {\bf 267}, 476 (1995). 

\bibitem{Diep} H.T.~Diep and H. ~Giacomini  in 
{\it Frustrated spin systems} edited by  H.T.~Diep,
 (World Scientific 2005).

\bibitem{Lois} D. ~Loison, in {\it Frustrated spin systems} 
edited by  H.T.~Diep, (World Scientific, 2005); cond-mat/0504326.  

\bibitem{Diehl}  H.W. ~Diehl, Acta Phys. Slov. {\bf 52}, 271 (2002);
 Pramana {\bf 64}, 803 (2005).


\bibitem{Malakis} A.~Malakis, P. ~Kalozoumis and N. ~Tyraskis, 
 Eur. Phys. J.  B {\bf 50}, 63 (2006).
 
\bibitem{Chit} G. Y. ~Chitov and C. ~Gros, Low Temp. Phys., 
{\bf 31}, 722 (2005).

\bibitem{Wid} B. Widom, J. Chem. Phys. {\bf 84}, 6943 (1986)

\bibitem{exp} Y.M. Vysochanskii and V.U. Slivka,
 Usp. Fiz. Nauk {\bf 162}, 139 (1992) (Sov. Phys. Usp. {\bf  35}, 123 (1992)); 
 F. S. Bates, W. Maurer, T. P. Lodge, M. F. Schulz, M. W. Matsen,
 K. Almdal and K. Mortensen,
Phys. Rev. Lett. {\bf 75}, 4429 (1995); 
 D. Schwahn, K. Mortensen, H. Frielinghaus and K. Almdal, Phys. Rev. Lett. 
{\bf 82}, 5056 (1999);
 M. Skarabot, R. Blinc, I. Musevic, A. Rastegar and T. Rasing, 
Phys. Rev. E {\bf 61}, 3961 (2000); 
I. Luk'yanchuk, A. Jorio and P. Saint-Gregoire, Phys. Rev. B {\bf 61},
 3147 (2000); C.C. Becerra, V. Bindilatti and N.F. Oliveira, 
Phys. Rev. B {\bf 62}, 8965 (2000); 
 H. Weitzel, H. Ehrenberg, C. Heid, H. Fuess, and  P. Burlet,
 Phys. Rev. B {\bf 62}, 12146 (2000).

\bibitem{Symanzik} K. ~Symanzik, Nucl. Phys. B {\bf 226}, 187 (1983);
  B. Berg, S. Meyer and I. Montvay, Nucl. Phys. B {\bf 235}, 149 (1984);
M. ~L\"uscher and P. Weisz, Comm. Math. Phys. {\bf 97}, 59 (1985).

\bibitem{Sheik} B. Sheikholeslami and R. Wohlert, 
Nucl. Phys. B {\bf 259}, 572 (1985).

\bibitem{Redner} S. ~Redner and H.E. ~Stanley, Phys. Rev. B .
{\bf 16}, 4901 (1977); J. Phys. C {\bf 10}, 4765 (1977).  


\bibitem{bcisiesse} P. Butera and M. Comi, 
 Phys. Rev. B {\bf 65}, 144431 (2002);
B {\bf 72}, 014442 (2005); J. Statist. Phys. {\bf 109}, 311 (2002). 

\bibitem{bper}  P. Butera and M. Pernici, Phys. Rev. B {\bf 76}, 092406 (2007)
 and arXiv:0806.1496 to be published.


\bibitem{guttmann} A. J. ~Guttmann, in {\it Phase Transitions and Critical
Phenomena}, edited by C.~Domb and J.~Lebowitz, 
(Academic, New York, 1989) Vol.\ 13. 

\bibitem{Oitmaa} M. ~Plischke and J. ~Oitmaa, 
Phys. Rev. B {\bf 19}, 487 (1979);
 J. ~Oitmaa, J. Phys. A {\bf 14}, 1159(1981);
 {\bf 18}, 365 (1985).

\bibitem{FerVe}M. J. Velgakis and M. ~Ferer, Phys. Rev. B {\bf 27}, 401 (1983).  


\bibitem{OitVe} J. ~Oitmaa and M. J. ~Velgakis, J. Phys. A: Math. Gen. {\bf 20}, 
1495 (1987).

\bibitem{Ferer} Z. ~Mo and M. ~Ferer, Phys. Rev. B {\bf 43}, 10890 (1991).

\bibitem{FisherPDA} M. E. ~Fisher in 
{\it Statistical Mechanics and Statistical methods in Theory and Applications}
 edited by U. Landman (Plenum, New York, 1977);
M. E. ~Fisher and R. M. Kerr, 
Phys. Rev. Lett. {\bf 39}, 667(1977). 

\bibitem{Alab} C. ~Alabiso and P. ~Butera, 
J. Math. Phys. {\bf 16}, 840 (1975).


\bibitem{Zinn81}J. ~Zinn-Justin, J. Physique (France) {\bf42}, 783 (1981).

\bibitem{Fisher}J. H. ~Chen, M. E. ~Fisher and B. G. ~Nickel, 
 Phys. Rev. Lett. {\bf 48}, 630 (1982);  M. E. ~Fisher and J. H. ~Chen, 
J. Physique(France)  {\bf 46}, 1645 (1985).

\bibitem{Bloete}H. W. J. ~Bl\"ote, E. ~Luijten and J. ~Heringa,
 J. Phys. A: Math. Gen. {\bf 28}, 6289 (1995)

\bibitem{xypisa} M. Campostrini, M. Hasenbusch, A. Pelissetto, P. Rossi
 and E. Vicari, Phys. Rev. B {\bf 63}, 214503 (2001).



\bibitem{Soe} A. ~Soehiane and J. ~Oitmaa, Mod. Phys. Lett. B {\bf 10}, 
609 (1997).
 
\bibitem{Clarizia} A. ~Clarizia, G.~ Cristofano, R. ~Musto, F. ~Nicodemi and
R. ~Pettorino, Phys. Lett. B {\bf 148}, 323 (1984). 


\bibitem{Binder}  D. P. ~Landau, J. Appl. Phys. {\bf 42}, 1284 (1971); 
D.P. ~Landau, Phys. Rev. B {\bf 21}, 1285 (1980);
 D. P. ~Landau and K. ~Binder, Phys. Rev. B {\bf 31}, 5946 (1985).

\bibitem{Sel78} W. ~Selke, Z. Phys.  B {\bf 29}, 133 (1978).

\bibitem{Sel79} W. Selke and M.E. Fisher, Phys. Rev. B{\bf 20}, 257 (1979).

\bibitem{Kaski} K. Kaski and W. Selke, Phys. Rev. B {\bf 31}, 3128 (1985). 

\bibitem{PHENK} M. Pleimling and M. Henkel, 
Phys. Rev. Lett. {\bf 87}, 125702 (2001).

\bibitem{Selke78} W. ~Selke, Solid St. Comm. {\bf 27}, 1417 (1978). 

\bibitem{Selke80}  W. ~Selke, J. Phys. C {\bf13}, L261, (1980).


\bibitem{Ferna} J. F. ~Fernandez, M. ~Puma and R. F. ~Angulo,
 Phys. Rev. B {\bf 44},  10057 (1991).


\bibitem{LoisonSimon} D. ~Loison and P. ~Simon, 
Phys. Rev. B {\bf 61}, 6114 (2000).

\bibitem{DiSh}  H. W. ~Diehl and M. ~Shpot,  Phys. Rev. B {\bf 62}, 
12338 (2000);  M. ~Shpot and H.W. ~Diehl, Nucl. Phys. B {\bf 612}, 340 (2001). 

\bibitem{lemer} M. M. Leite, Phys. Rev. B {\bf 67}, 104415 (2003);
C.  Mergulh\~ao, Jr. and C. E. I. Carneiro, Phys. Rev. B {\bf 58}, 6047 (1998);
 B {\bf 59}, 13954 (1999); P. R. S. Carvalho and M. M. Leite, arXiv:0704.3208.



\bibitem{Berv} C. ~Bervillier, Phys. Lett. A {\bf 331}, 110 (2004).

\bibitem{Garel} T. Garel and P. Pfeuty, J. Phys. C: Solid State Phys., 
 {\bf 9}, L245 (1976).


\bibitem{ahamuk} A. Aharony and D. Mukamel, J. Phys. C {\bf 13}, L255 (1980).

\bibitem{Shpot} M. A. ~Shpot, Yu. M. ~Pis'mak and H. W. ~Diehl, 
 J. Phys. Condens. Matter {\bf17}, S1947 (2005).

\bibitem{Hor}R.M. ~Hornreich, M. ~Luban and S. ~Shtrikmann,
 Phys. Lett. A {\bf 55}, 269 (1975).

\bibitem{HB}
 D. Mukamel, J. Phys. A {\bf 10}, l249 (1977); 
 A. D. Bruce and R.M. ~Hornreich, J. Phys. A {\bf 11}, 595 (1978).
  
\bibitem{bcoen} P. ~Butera, and M.~Comi, Phys. Rev. B {\bf 56}, 8212 (1997). 

\bibitem{LiFi}A. J. Liu and M. E. Fisher, 
 J. Stat. Phys.  {\bf 58}, 431 (1990).
 
\bibitem{cpr} D. L. Hunter and G. A. Baker, Phys. Rev. B {\bf 7}, 3346 (1972).

\bibitem{BCsi} P. ~Butera, and M.~Comi, Phys. Rev. B {\bf 60}, 6749 (1999).  


\bibitem{Hase}M. Hasenbusch and T. T\"or\"ok, J. Phys. A {\bf 32}, 6361 (1999).

\bibitem{buro} E. Burovski, J. Machta, N. Prokof'ev and B. Svistunov,
Phys. Rev. B {\bf 74}, 132502 (2006).


\bibitem{beach} K. S. D. Beach, cond-mat/0507541.

\bibitem{Lipa} J.A. Lipa, J. A. Nissen,D. A. Stricker, D. R. Swanson and
 T. C. P. Chui, Phys. Rev. B {\bf 68}, 174518 (2003).

\bibitem{nasser} I. Nasser and R. Folk, Phys. Rev. B {\bf 52}, 15799 (1995).


\bibitem{Diep2} V. Thanh Ngo and H, T. Diep, J. App. Phys. 
 {\bf 103}, 07C712 (2008).

\bibitem{Tiss} M. Tissier, B. Delamotte, and D. Mouhanna, 
Phys. Rev. B {\bf 67}, 134422 (2003).


\bibitem{SDEM} P. ~Butera, M. ~Comi and G. ~Marchesini, Phys. Rev. B {\bf 33},
  4725 (1986).

\bibitem{SDEA} P. ~Butera, M. ~Comi and G. ~Marchesini, 
Nucl. Phys. B {\bf300}, 1 (1988);
 P. ~Butera, M. ~Comi and G. ~Marchesini, Phys. Rev. B {\bf 40}, 534 (1989);
P. ~Butera, and M. ~Comi, {\it ibid.} B {\bf 46}, 11141 (1992);
 B {\bf 50}, 3052 (1994).


\bibitem{SDED}  P. ~Butera, R. ~Cabassi, M. ~Comi and G. ~Marchesini, 
Comput. Phys. Commun. {\bf 44}, 143 (1987);
P. ~Butera, and M. ~Comi, Ann. Comb. {\bf 3}, 277 (1999).

\bibitem{Rednerpc} Private communication from Prof. S. ~Redner.

\bibitem{bc} P. ~Butera, and M.~Comi, Phys. Rev. B {\bf 54}, 15828 (1996).


\end{thebibliography}
\end{document}